\documentclass[amsmath,amssymb,amsbsy,prl,twocolumn,preprintnumbers,showpacs]{revtex4-2}
\usepackage{graphics,color}
\usepackage{dcolumn}
\usepackage{bm}
\usepackage{braket}
\usepackage{mathtools}
\usepackage{ulem}
\usepackage[breaklinks,colorlinks=true,linkcolor=blue,urlcolor=blue,citecolor=cyan]{hyperref}

\begin{document}
\title{Detecting Bulk Topology of Quadrupolar Phase from Quench Dynamics}
\author{Tomonari Mizoguchi}
\affiliation{Department of Physics, University of Tsukuba, Tsukuba, Ibaraki 305-8571, Japan}
\email{mizoguchi@rhodia.ph.tsukuba.ac.jp}
\author{Yoshihito Kuno}
\affiliation{Department of Physics, University of Tsukuba, Tsukuba, Ibaraki 305-8571, Japan}
\author{Yasuhiro Hatsugai}
\affiliation{Department of Physics, University of Tsukuba, Tsukuba, Ibaraki 305-8571, Japan}

\begin{abstract}
Direct measurement of a bulk topological observable in topological phase of matter has been a long-standing issue. 
Recently, detection of bulk topology through quench dynamics has attracted growing interests.
Here, we propose that topological characters of a quantum quadrupole insulator 
can be read out by quench dynamics.
Specifically, we introduce a quantity, a quadrupole moment weighted by the eigenvalues of the chiral operator, 
which takes zero for the trivial phase and finite for the quadrupolar topological phase. 
By utilizing an efficient numerical method to track the unitary time evolution, we elucidate that the quantity we propose indeed serves 
as an indicator of topological character for both noninteracting and interacting cases. 
The robustness against disorders is also demonstrated. 
\end{abstract}
\maketitle

\textit{Introduction.---}
Understanding topological aspects of quantum matters has been one of the central issues in modern condensed matter physics~\cite{PhysRevLett.49.405,PhysRevLett.61.2015}.
Discovery of topological insulators (TIs)~\cite{PhysRevLett.95.146802,PhysRevLett.95.226801, Bernevig1757,RevModPhys.82.3045,RevModPhys.83.1057} is highlighted as one of the most
prominent steps that makes the roles of topology manifest. 
Specifically, it was found that topological natures of Bloch electrons characterized 
by topological invariants result in boundary modes robust against perturbations~\cite{PhysRevB.25.2185}. 
This relation between bulk topology and boundary modes is called bulk-boundary correspondence (BBC), and it has served as a central notion in studies on topological materials~\cite{PhysRevLett.71.3697,PhysRevB.48.11851}.

BBC also ties topologically protected boundary modes with quantized responses to external fields, which is another 
characteristic of TIs.
A representative example is the quantum Hall effect where the number of edge modes corresponds to 
the Hall conductance~\cite{PhysRevB.25.2185,PhysRevLett.71.3697,PhysRevB.48.11851}. 
Another example is the quantization of an electric dipole moment attributed to the quantized Berry's phase 
of Bloch electrons in one dimension~\cite{PhysRevLett.62.2747,PhysRevB.47.1651,PhysRevB.48.4442,RevModPhys.66.899,Ryu2002,PhysRevB.69.085106,PhysRevB.78.195424,HATSUGAI20091061,PhysRevB.86.115112,PhysRevX.8.021065}. 
From the viewpoint of BBC, this is attributed to the boundary states localized at the ends of one-dimensional systems. 
Recently, this topological viewpoint of an electric dipole 
is further extended~\cite{0295-5075-95-2-20003,PhysRevB.90.085132} to higher-rank multipole moments~\cite{PhysRevB.96.245115,Benalcazar61,PhysRevB.100.245133,PhysRevB.100.245134,PhysRevB.100.245135,Watanabe2020} (e.g., quadrupole and octapole) in two or higher dimensions,
that are attributed to the boundary states localized at the corners.
Such a topological phase of matter hosting boundary modes with 
codimension greater than one is 
nowadays established as a higher-order topological phase,
and large amount of theoretical~\cite{PhysRevB.95.165443,Hayashi2018,Xu2017,Schindlereaat0346,PhysRevLett.120.026801,PhysRevB.97.205136,PhysRevB.97.241405,PhysRevB.98.035147,PhysRevB.98.235102,PhysRevB.99.041301,PhysRevB.99.085127,PhysRevB.99.085406,PhysRevB.99.235132,PhysRevB.99.245151,PhysRevLett.123.196402,PhysRevB.100.235302,PhysRevResearch.2.012009,PhysRevB.101.085137,PhysRevB.101.115140} and experimental~\cite{Schindler2018,Garcia2018,Imhof2018,PhysRevB.98.205147,Peterson2018,Noh2018,Mittal2019,Ota:19,ElHassan2019,NatMaterXue2019,Ni2019,Zhang2019,Zhang2019_2,Kempkes2019} efforts have been devoted to understanding and realizing this phase.

\begin{figure}[b]
\begin{center}
\includegraphics[clip,width = 0.98\linewidth]{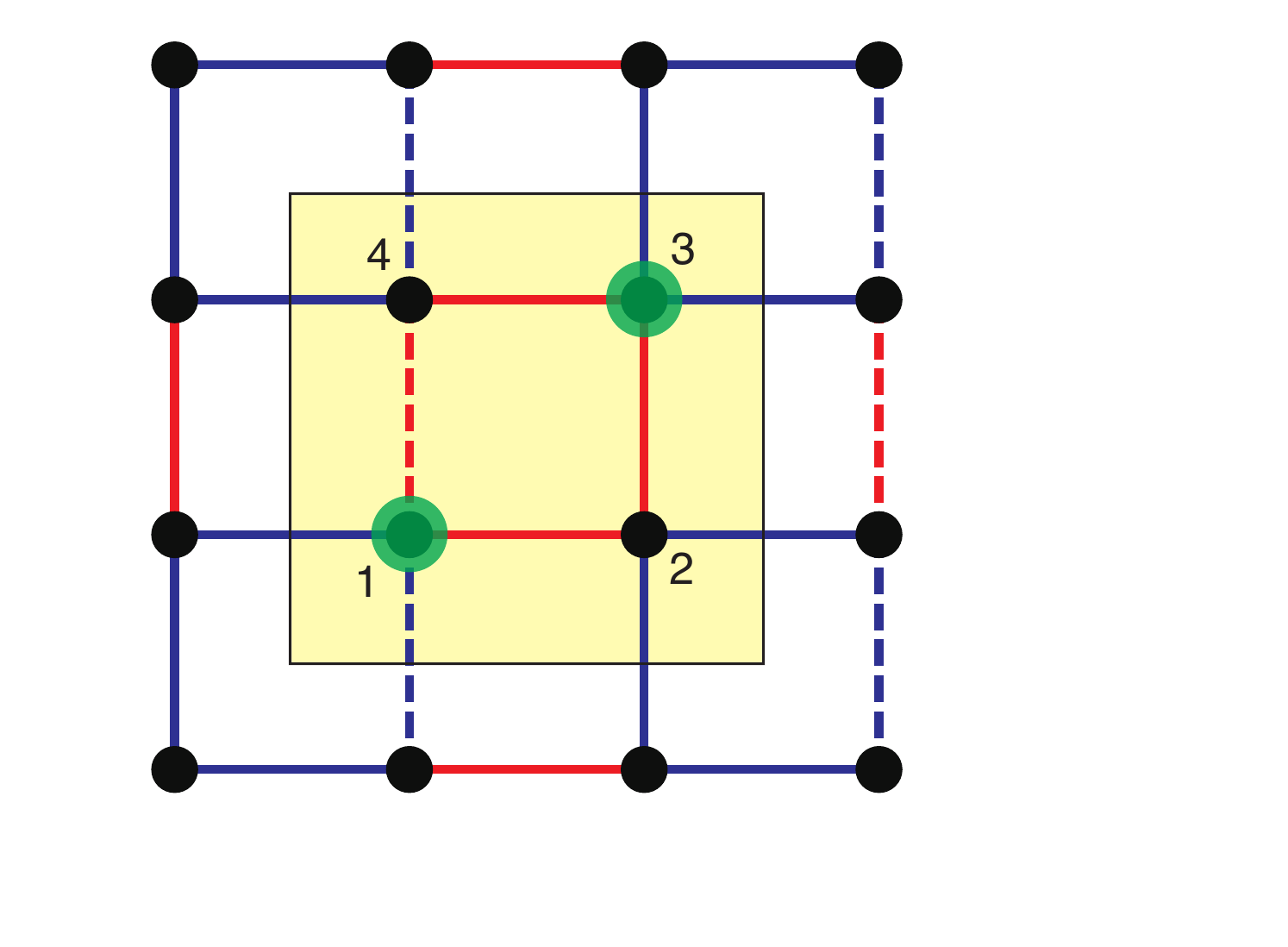}
\vspace{-10pt}
\caption{Schematics of the BBH model.
The unit cell is indicated by a yellow shade,
and green circles schematically represent the initial positions of the two particles.
Note that $\pi$ flux threads each square plaquette.}
  \label{fig:model}
 \end{center}
 \vspace{-10pt}
\end{figure}
It had been a common belief that topological invariants
themselves are not observed from featureless gapped ground states, but characteristic boundary modes enable us to observe them. 
However, recently, an approach to directly access bulk topological natures was proposed, 
that is, dynamical aspects of topological phases~\cite{PhysRevB.53.7010,RevModPhys.82.1959,Atala2013,Mazza_2015,Meier2016,PhysRevLett.118.185701,Cardano2017,Meiereaat3406,Maffei_2018,PhysRevB.97.060304,PhysRevLett.121.250601,Zhang2018,PhysRevA.99.053606,PhysRevLett.124.160402,Longhi:19,Qiu2019,Wang2019,Wang2019_2,Haller2020,Ji2020}.
Such attempts are essential because they are beyond the aforementioned common belief on topological phases.
In the literature, the semiclassical approach of wave-packet dynamics has successfully 
illustrated the role of Berry curvature in transport phenomena~\cite{PhysRevB.53.7010,RevModPhys.82.1959}.
Another direction, on which we focus in this Letter, is considering 
quench dynamics from completely localized initial states.
At single-particle level, 
i.e., without the Fermi sea of the filled bands,
the information of Bloch bands in an entire Brillouin zone can be embedded 
by setting a spatially localized initial state; this is attributed to the fact that the localized states can be expressed 
as a linear combination of all the states in the Fourier space.
As such, topological invariants can be extracted from the time-dependent quantities. 
For instance, for chiral-symmetric one-dimensional systems, 
the quantity called
the mean chiral displacement (MCD), which is the polarization weighted by the eigenvalue of the chiral operator, successfully extracts the topological winding number of the system,
and thus distinguishes the topological states from the trivial states~\cite{Cardano2017,Meiereaat3406,Maffei_2018,Haller2020}.
Moreover, measurements of such quantities are experimentally feasible in 
various setups, e.g., discrete quantum walk in a photonic system based on the orbital angular momentum of a light beam~\cite{Cardano2017}. 

Considering the findings listed above, one is naturally tempted to ask the following questions: 
(i) Can we apply the measurement of topological invariants through dynamics to higher-order topological (or quadrupolar) phases?
(ii) If so, can it be also applicable to interacting systems and/or disordered ones?
In this Letter, to address these issues, 
we investigate two-particle dynamics of the interacting Benalcazar-Bernevig-Hughes (BBH) model~\cite{PhysRevB.96.245115,Benalcazar61}. 
We heuristically find a quantity whose long time average can characterize the topology. This quantity is a modified bulk quadrupole moment, which is reminiscent of the MCD in one-dimensional systems. 
Therefore, this quantity is also experimentally measurable.
By using a numerically efficient algorithm of tracing the unitary time evolution of the two-particle wave function (one may increase the number of particles in principle),
we elucidate that the quantity introduced here characterizes the topological nature of the BBH model, for both noninteracting and interacting cases. 
Furthermore, the characteristic behavior of this quantity is robust against moderate strength of disorders,
indicating the feasibility of experimental measurements in realistic setups that are not completely clean. 

Hereafter, we set $\hbar = 1$. 

\textit{Model and method.---} We consider the model proposed in Refs.~\cite{PhysRevB.96.245115,Benalcazar61},
incorporating an interaction and a disorder. 
The Hamiltonian reads
$H = H _0 + H_{\rm int} + H_{\rm rand}$,
where
$H_0 = \sum_{\langle i,j \rangle} t_{i,j} a^{\dagger}_i a_{j} + (\mathrm{H.c.})$,
$H_{\rm int} = V \sum_{\langle i,j \rangle} n_{i} n_j$,
and
$H_{\rm rand} = \sum_{i} w_i n_{i}$.
Here $a$ and $a^\dagger$ denote, respectively, the annihilation and creation operators of spinless fermions, 
and $i$ denotes the sites on a square lattice specified by a pair of indices $\bm{r}$ and $\alpha = 1,2,3,4$,
where $\bm{r} = (r_x, r_y)$ is the position of the unit cell, and $\alpha$ labels the sublattice (Fig.~\ref{fig:model}).
$n_i:= a^\dagger_i a_i$ is the density operator.   
The symbol $\langle, \rangle$ represents the nearest-neighbor pairs of sites.
The transfer integral $t_{i,j}$ is indicated in Fig.~\ref{fig:model}; there are two parameters, $t_a$ and $t_b$.
We note that $H_0$ preserves the chiral symmetry, such that $\hat{\Gamma} H_0 \hat{\Gamma} = -H_0$
with $\hat{\Gamma} =e^{i \pi \sum_{\bm{r}} \left(n_{\bm{r},2} + n_{ \bm{r},4} \right) } $.
In addition to $H_0$, we consider two terms $H_{\rm int}$ and $H_{\rm rand}$.
Here, $V$ denotes the strength of the intersite interaction
and $w_i$ is the strength of the disorder potential, chosen randomly in $\left[-\frac{W}{2}, \frac{W}{2} \right]$.

The topological properties of the hopping term $H_0$ has been well investigated in the literature.
For $|t_a| \neq |t_b|$, the system is gapped at the half-filling. 
The half-filled ground state is topologically trivial (nontrivial) when $|t_a| > |t_b|$ ($|t_a| < |t_b|$).
The topological nature can be captured by topological invariants such as the nested Wilson loop~\cite{PhysRevB.96.245115,Benalcazar61}, 
the quadrupole moment~\cite{PhysRevB.100.245133,PhysRevB.100.245134,PhysRevB.100.245135}, 
the entanglement-related quantities~\cite{PhysRevB.98.035147,Wang_2018,PhysRevB.101.115140}, 
and the Berry phase~\cite{PhysRevResearch.2.012009}.
Furthermore, nontrivial topology results in the emergence of the corner states, which is characteristics of the higher-order topological phases. 
The aim of this study is to extract the topological nature without relying on the corner states.

The quench dynamics of the system can be dictated by
the unitary time evolution of the many-body wave function, $\ket{\Psi(t)} = e^{-i H t} \ket{\Psi(0)}$. 
To obtain $\ket{\Psi(t)}$ numerically, we approximate $e^{-i H t}$ as follows.
First, we discretize the time as $t_l =  l \Delta \tau$, with $\Delta \tau$ being small time step 
(compared with the hopping parameters); we set $\Delta \tau = 0.01$ in the present work.
Then, we have $e^{-i H t_l} \sim \left( e^{-i H \Delta \tau}\right)^l$.
The remaining task is to approximate $e^{-i H \Delta \tau}$. To this end, 
we employ
the fourth-order Suzuki-Trotter decomposition~\cite{doi:10.1142/S0217979201004885,SUZUKI1990319}, 
namely,
$e^{-i \Delta \tau H} = S(-i p \Delta \tau) S(-i (1-2p) \Delta \tau) S(-i p \Delta \tau)$,
where $p:= \left(2-2^{1/3} \right)^{-1}$ 
and $S(x) = e^{x \frac{H_1}{2}} \cdots e^{x \frac{H_{q-1}}{2} } e^{x H_q} e^{x \frac{H_{q-1}}{2} } \cdots e^{x \frac{H_1}{2}}.$
Note that, in defining $S(x)$, we divide the Hamiltonian $H$ into $q$ pieces, $H = H_1 + \cdots +H_q$, which do not necessarily commute each other.
Here, we set $q = 5$, and we show the explicit forms of $H_1$-$H_5$ in Supplemental Material~\cite{SM}.
The Suzuki-Trotter decomposition of $e^{-i H t}$ largely reduces computational costs. 
Hence, we can access long time dynamics with relatively large system size 
in short computational time,
compared with other methods such as exact diagonalization. 

\begin{figure}[t]
\begin{center}
\includegraphics[clip,width = 0.98\linewidth]{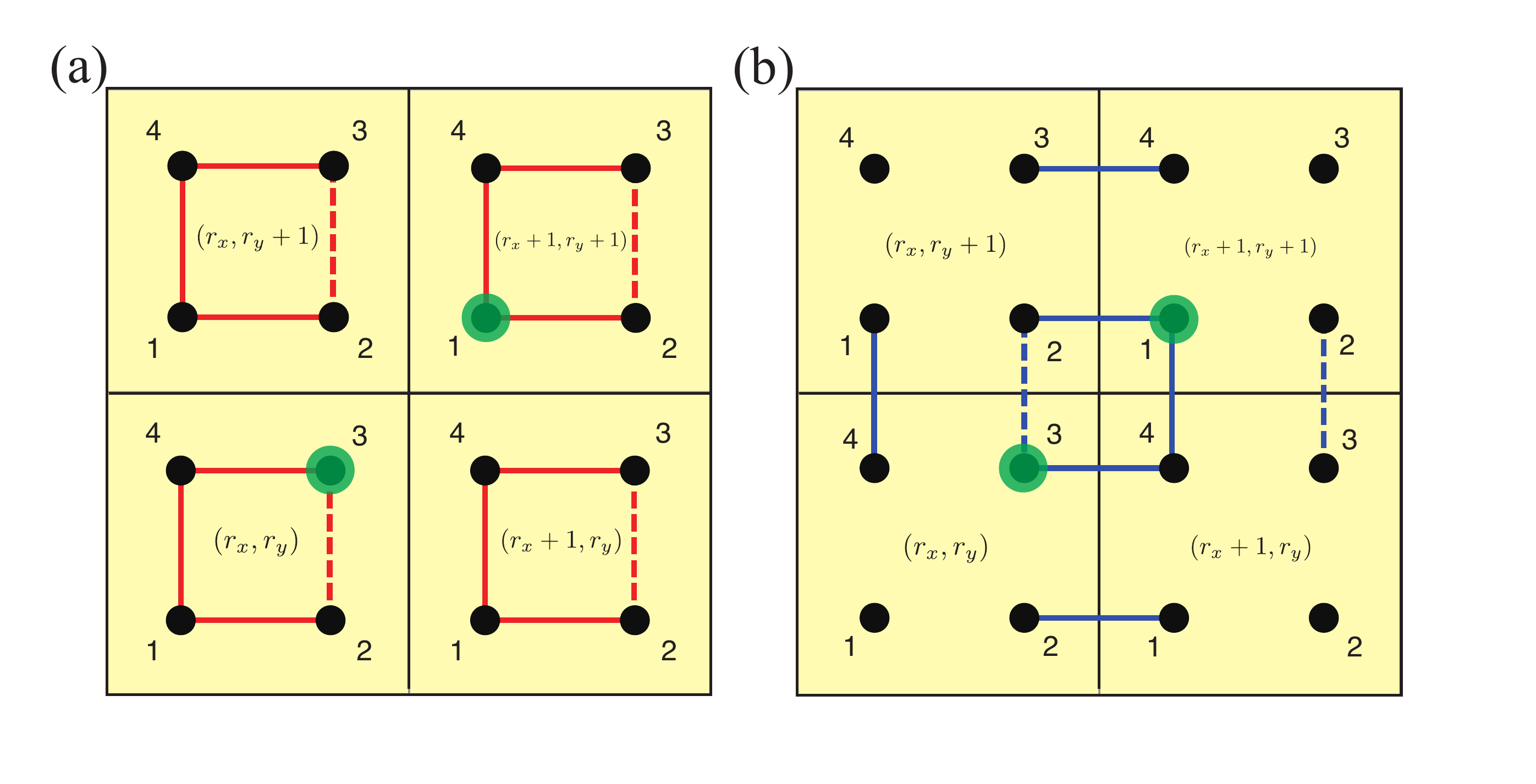}
\vspace{-10pt}
\caption{Schematic figures of two decoupled limits. 
The panel (a) corresponds to $t_b =0$, i.e., the topologically trivial case,
and (b) corresponds to $t_a =0$, i.e., the topologically nontrivial case.}
\label{fig:CQM}
\end{center}
 \vspace{-10pt}
\end{figure}
\begin{figure*}[t]
\begin{center}
\includegraphics[clip,width = \linewidth]{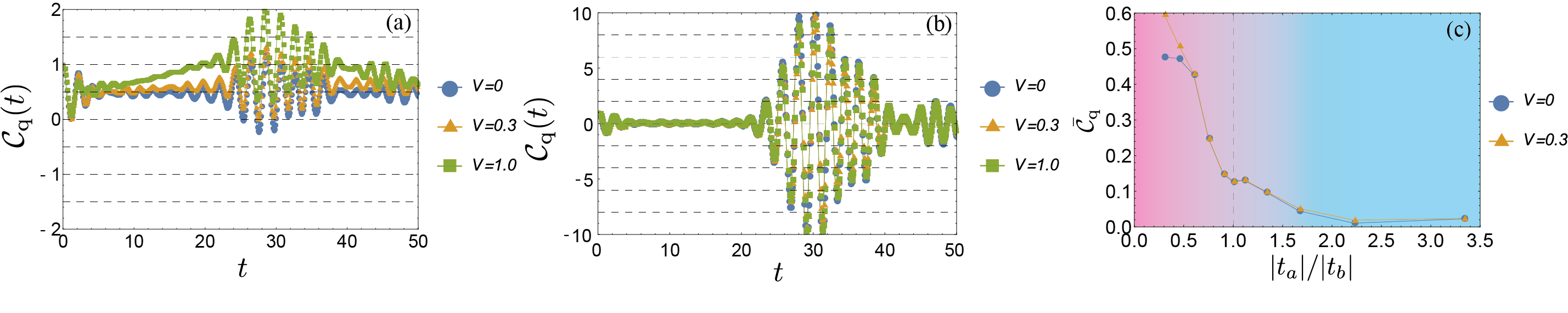}
\vspace{-10pt}
\caption{The MCQM for the clean system (i.e., $W=0$) with (a) $t_a=-0.3$, $t_b = -1.0$ and 
(b) $t_a=-1.0$, $t_b = -0.3$.
(c) The time-averaged MCQM, $\bar{\mathcal{C}}_{\rm q}$, as a function of $|t_a|/|t_b|$.
The average is taken over $t\in [0,50]$, and the parameters in the actual simulations are set such that max$\{ |t_a|, |t_b |\} = 1$.} 
  \label{fig:result_1}
 \end{center}
 \vspace{-10pt}
\end{figure*}
\textit{Mean chiral quadrupole moment.---} 
The main proposal of this Letter is the introduction of a quantity characterizing the topological nature of the quadrupolar phase, which may be termed the mean chiral quadrupole moment (MCQM):
\begin{eqnarray}
\mathcal{C}_{\rm q} (t) = \bra{\Psi(t)} \mathcal{Q} \ket{\Psi(t)}, 
\end{eqnarray}
where
\begin{eqnarray}
\mathcal{Q}  = \sum_{\bm{r},\alpha} r_x r_y \Gamma_\alpha n_{\bm{r},\alpha}. \label{eq:Q}
\end{eqnarray}
Here $\Gamma_\alpha$ is the eigenvalue of the chiral operator; it takes $1$ for $\alpha = 1,3$ and $-1$ for $2,4$. 
Note that, to make the MCQM well defined, we have to fix the labeling of the unit cells and sublattices in the beginning, since this quantity depends on the choice of the frame.
Except for $\Gamma_\alpha$, $\mathcal{Q}$ follows the conventional definition of the quadrupole operator 
under the open boundary condition~\cite{PhysRevB.96.245115,Benalcazar61,PhysRevB.100.245133,PhysRevB.100.245134,PhysRevB.100.245135,Watanabe2020}.
This quantity can be regarded as an extension of the MCD which dictates the winding number of one-dimensional TIs in classes AIII and BDI in the topological classification, having even number of bands~\cite{Cardano2017,Meiereaat3406,Maffei_2018,Haller2020}.
Note that, in actual experiments, all we need to measure is the site-resolved particle density.
This guarantees accessibility of this quantity if spatial resolution of experimental setup is sufficiently fine.
It is also noteworthy that this quantity is sensitive to the choice of the initial state.
In the present study, we choose the initial state such that two particles are localized 
at two diagonal sites on the inter-unit-cell plaquette located at the middle of the system, as schematically depicted as green circles in Fig.~\ref{fig:model}.
In Supplemental Material, we show the numerical data for a different choice of initial state, namely, two particles are localized 
at two diagonal sites on the intra-unit-cell plaquette, where we see that the failure of the distinction between topological and trivial cases.

How does the MCQM extract the topological nature of the quadrupole insulators?
To see this, we present an intuitive understanding of the implication of the MCQM, namely,
the argument of the decoupled four-site cluster limit.
In the prior works~\cite{doi:10.1143/JPSJ.75.123601,Hatsugai_2007,0295-5075-95-2-20003,PhysRevResearch.2.012009,PhysRevLett.123.196402,doi:10.7566/JPSJ.88.104703},
it was found that this argument is essential for understanding the ground-state properties of the insulating state, 
since the ground state is adiabatically connected to this limit and 
topological properties of gapped ground states are unchanged under the change of model parameters unless the excitation gap is closed. 
Regarding the dynamical properties, for which the information of all the eigenstates matters, the notion of adiabatic connection does not hold straightforwardly, but it still gives a useful insight. 
In fact, such an argument works in one-dimensional systems as well~\cite{SM}.

For the decoupled limits, the particles are confined in the plaquette on which the particle is initially located,
thus unitary time evolution can exactly be tracked by solving the four-site problem.
Thus, in these limits, the exact form of $\mathcal{C}_q (t)$ is accessible.
For the details of the calculations, see Supplemental Material~\cite{SM}.
For the present choice of the initial state, 
we find that, for the trivial limit, i.e., $t_b = 0$ [Fig.~\ref{fig:CQM}(a)],
one has 
\begin{eqnarray}
\bar{\mathcal{C}}_{\rm q}= 0,\label{eq:result1}
\end{eqnarray}
where $\bar{\mathcal{C}}_{\rm q}$ stands for the long time average of $\mathcal{C}_q (t)$.
Meanwhile, for the nontrivial limit i.e., $t_a = 0$ [Fig.~\ref{fig:CQM}(b)],
one has 
\begin{eqnarray}
\bar{\mathcal{C}}_{\rm q} =\frac{1}{2}. \label{eq:result2}
\end{eqnarray}
Equation~(\ref{eq:result2}) indicates that nonvanishing value of $\bar{\mathcal{C}}_{\rm q}$ 
under the proper choice of the initial state reflects the presence of  the nontrivial topology in the bulk.
It should be noted that the difference between the trivial limit and the the nontrivial limit is 
whether the plaquettes with finite hoppings are intra-unit-cell ones or an inter-unit-cell ones.
In fact, even away from the limiting cases, 
assigning larger hoppings on inter-unit-cell plaquettes than the intra-unit-cell ones is essential to obtain the finite value of $\bar{\mathcal{C}}_{\rm q}$,
as we will show later.

We briefly remark the role of $\Gamma_\alpha$.
In fact, the similar factor is included in the MCD for one-dimensional systems~\cite{Meiereaat3406,Maffei_2018}. 
In that case, its role is to make the contributions from the negative-energy bands and those from the positive-energy bands additive;
otherwise they cancel each other.
This fact also implies that the MCD is adaptable to chiral symmetric systems with an even number of topological bands.
In fact, $\Gamma_\alpha$ in the MCQM is incorporated in the same spirit, but in a rather heuristic manner.
Nevertheless, it is indeed essential 
so that the MCQM serves as a topological marker, as clarified in the decoupled cluster argument~\cite{SM}.
Moreover, the topological characterization is valid even in the presence of the chiral-symmetry-breaking term, 
$H_{\rm rand}$, as we will show later.

\textit{Numerical demonstration.---} 
We now demonstrate the validity of the MCQM for topological characterization.
In Figs.~\ref{fig:result_1}(a) and \ref{fig:result_1}(b), 
we plot $\mathcal{C}_{\rm q}(t)$ to $t = 50$ for topological and trivial cases respectively,
for the clean systems (i.e., $W=0$). 
Here the numerical computations are carried out for $20 \times 20$-site (i.e., $10 \times 10$-unit cell) systems under the open boundary condition.
At the initial state, two particles are localized at 
the sublattice 1 at the unit cell $\left(0,0 \right)$ and the sublattice 3 at the unit cell $\left(-1,-1\right)$. 
We see in these figures that, for the topological case with weak interaction ($V=0, 0.3$), $\mathcal{C}_{\rm q}(t)$ oscillates around $1/2$ as expected, whereas it oscillates around $0$ for the trivial case. 
Therefore, the long time average of $\mathcal{C}_{\rm q}(t)$ indeed can be used to extract the topological character of 
this model.

We also remark the boundary effects. 
In fact, the particles are initially located near the center of the system, 
and they reach the boundary at $t \sim 20$.
Although the amplitude of the oscillation of the MCQM increases after reaching the boundary,
the center of the oscillation is still unchanged, manifesting the robustness of $\bar{\mathcal{C}}_{\rm q}$ against the boundary effects.
Also, the fact that the finite value of MCQM in the topological case is obtained before reaching the boundary 
indicates that the finite MCQM is not attributed to the corner states, and thus this is indeed the bulk property. 

It can also be found in Fig.~\ref{fig:result_1}(a) that the role of the interaction becomes manifest even for moderate strength of the interaction ($V=1.0$).
In the topological case, the MCQM deviates from the noninteracting case, namely, 
the MCQM exhibits gradual increase (decrease) to $t \lesssim 30$ ($t \gtrsim 30$). 
This value of $V$ is smaller than the band gap at the half-filling.
This result indicates the essential difference between the dynamical properties and the ground-state properties 
at the half-filling, because the latter is stable against interactions as far as the excitation gap is not closed~\cite{PhysRevResearch.2.012009}. 
We also note that, in the trivial case, the MCQM seems to be rather insensitive to the interaction strength. 
\begin{figure}[t]
\begin{center}
\includegraphics[clip,width = 0.9\linewidth]{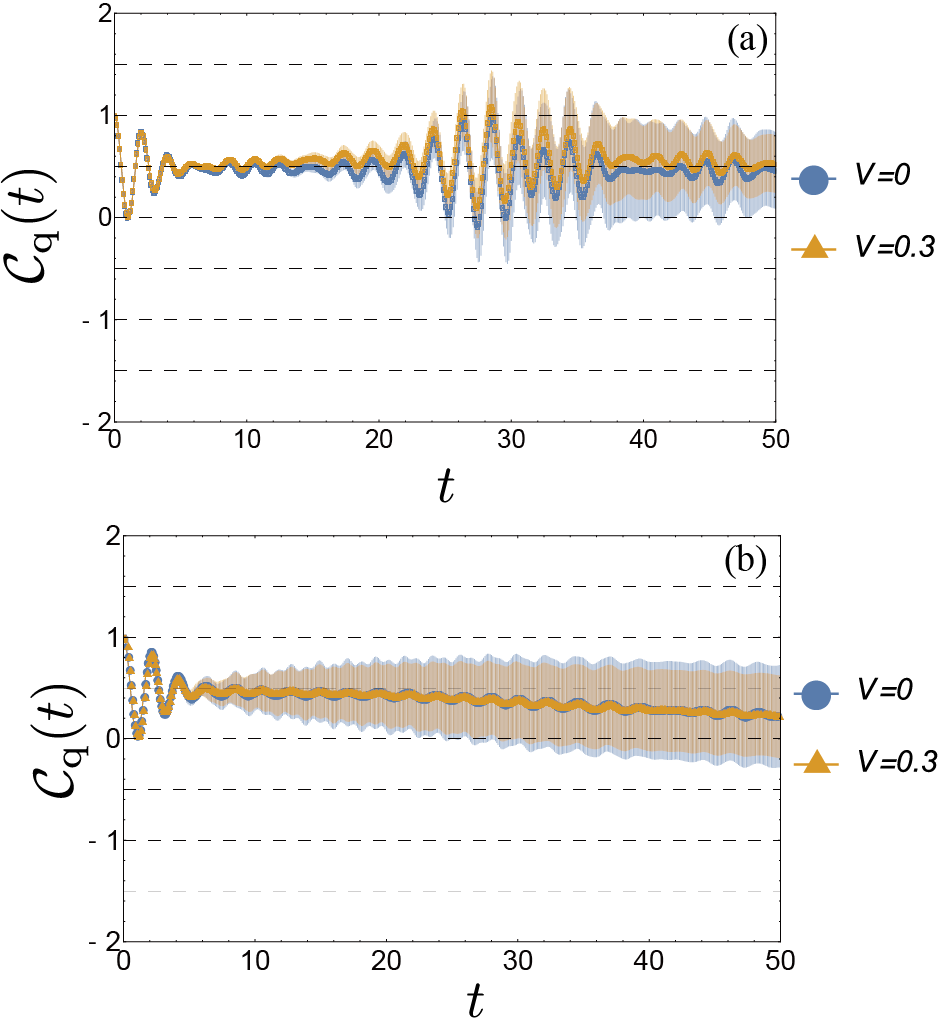}
\vspace{-10pt}
\caption{The MCQM for the disordered system with (a) $t_a=-0.3$, $t_b = -1.0$, $W=0.2$
and (b) $t_a=-0.3$, $t_b = -1.0$, $W=1.0$. The error bars are represented by the shades.}
  \label{fig:result_2}
 \end{center}
 \vspace{-10pt}
\end{figure}

It should also be noted that the ``topological transition" of the dynamical properties
is not as sharp as that for the ground state.
To show this, we plot $\bar{\mathcal{C}}_{\rm q} (t)$ as a function of $|t_a|/|t_b|$ in Fig.~\ref{fig:result_1}(c).
We see that the dependence on $|t_a|/|t_b|$ of $\bar{\mathcal{C}}_{\rm q}$ is smooth, rather than a steep jump; 
this is another indication of the difference between the dynamical properties and the ground-state properties, 
as the latter is characterized by the jump of the quantized topological number. 
Note that, at the critical point ($t_a = t_b$), the behavior of $\mathcal{C}_{\rm q} (t)$ is qualitatively different from that deep inside the topological or trivial cases, namely, $\mathcal{C}_{\rm q} (t)$ does not oscillate around a certain value; 
see Supplemental Material~\cite{SM}.
This behavior will serve as a useful hallmark indicating that the system is near the critical point.

\textit{Robustness against disorders.---} 
We further study the effects of disorder potentials, to test the robustness of the MCQM. 
In Fig.~\ref{fig:result_2}, we plot $\mathcal{C}_{\rm q}(t)$ for the topological case with weak ($W=0.2$) and moderate ($W=1.0$) disorders. 
Here the average is taken over 432 configurations of the random disorder potential. 
We see that $\mathcal{C}_{\rm q}(t)$ remains to oscillate around $1/2$ for a weak disorder case.
In particular, for $t \lesssim 20$ where the particles do not reach the boundary, the error bars due to the disorder average are very small. 
Even for the moderate disorder case, the short-time behavior (i.e., $t\lesssim5$) is almost unaffected by the disorders.
However, after the long time ($t \gtrsim 20$), the $\mathcal{C}_{\rm q}(t)$ starts to decrease gradually and deviates from $1/2$.
These results manifest the robustness of the MCQM against disorders, which indicates that this is a measurable quantity in experiments 
for moderately clean samples.  

\textit{Summary.---}
We have proposed how to extract the topological character 
of the quadrupolar phase by the quench dynamics. 
Specifically, we introduce the MCQM as a marker of a topological character.
Although the initial state in the present setup is localized, the system is translationally invariant without any boundaries.
The numerical results on the two-particle BBH model indicate that 
the MCQM indeed captures the topological nature for weakly interacting and moderately clean systems.
It has also been clarified that there are essential differences 
between the ground-state topological properties at the half-filling and the MCQM,
with respect to the stability against the interactions and the sharpness of the topological transition.
This is attributed to the fact that the former is protected by the finite excitation gap,
whereas the latter is affected by the information of all the eigenvalues and eigenvectors. 

It is worth pointing out that our method of tracking two-particle dynamics is also applicable to bosonic systems.
We find that the results are qualitatively the same as those for fermions~\cite{SM}.
This may indicate that each particle contributes to the MCQM additively as far as the few-body systems are concerned and it holds regardless of whether the system is bosonic or fermionic.
Considering this fact, the present method can cover a wide range of experimental setups,
including fermionic and bosonic ultracold atoms under the optical lattice, photonic crystals, and discrete quantum walks. 
Meanwhile, insensitivity for the particle statistics may not hold for the many-bony systems and studying such cases is an interesting future problem~\cite{Haller2020}.
We hope our proposal opens up a way to understanding novel aspects of the quadrupolar phase.

\acknowledgments
This work is supported by the JSPS KAKENHI, 
Grants No.~JP17H06138, NO.~JP20K14371 (T. M.), Japan,
and JST CREST, Grant No. JPMJCR19T1, Japan.
Parts of numerical calculations were carried out on the 
Supercomputer Center at Institute for Solid State Physics, University of Tokyo.

\bibliographystyle{apsrev4-2}
\bibliography{HOTI}

\begin{thebibliography}{88}%
\makeatletter
\providecommand \@ifxundefined [1]{%
 \@ifx{#1\undefined}
}%
\providecommand \@ifnum [1]{%
 \ifnum #1\expandafter \@firstoftwo
 \else \expandafter \@secondoftwo
 \fi
}%
\providecommand \@ifx [1]{%
 \ifx #1\expandafter \@firstoftwo
 \else \expandafter \@secondoftwo
 \fi
}%
\providecommand \natexlab [1]{#1}%
\providecommand \enquote  [1]{``#1''}%
\providecommand \bibnamefont  [1]{#1}%
\providecommand \bibfnamefont [1]{#1}%
\providecommand \citenamefont [1]{#1}%
\providecommand \href@noop [0]{\@secondoftwo}%
\providecommand \href [0]{\begingroup \@sanitize@url \@href}%
\providecommand \@href[1]{\@@startlink{#1}\@@href}%
\providecommand \@@href[1]{\endgroup#1\@@endlink}%
\providecommand \@sanitize@url [0]{\catcode `\\12\catcode `\$12\catcode
  `\&12\catcode `\#12\catcode `\^12\catcode `\_12\catcode `\%12\relax}%
\providecommand \@@startlink[1]{}%
\providecommand \@@endlink[0]{}%
\providecommand \url  [0]{\begingroup\@sanitize@url \@url }%
\providecommand \@url [1]{\endgroup\@href {#1}{\urlprefix }}%
\providecommand \urlprefix  [0]{URL }%
\providecommand \Eprint [0]{\href }%
\providecommand \doibase [0]{https://doi.org/}%
\providecommand \selectlanguage [0]{\@gobble}%
\providecommand \bibinfo  [0]{\@secondoftwo}%
\providecommand \bibfield  [0]{\@secondoftwo}%
\providecommand \translation [1]{[#1]}%
\providecommand \BibitemOpen [0]{}%
\providecommand \bibitemStop [0]{}%
\providecommand \bibitemNoStop [0]{.\EOS\space}%
\providecommand \EOS [0]{\spacefactor3000\relax}%
\providecommand \BibitemShut  [1]{\csname bibitem#1\endcsname}%
\let\auto@bib@innerbib\@empty
\bibitem [{\citenamefont {Thouless}\ \emph {et~al.}(1982)\citenamefont
  {Thouless}, \citenamefont {Kohmoto}, \citenamefont {Nightingale},\ and\
  \citenamefont {den Nijs}}]{PhysRevLett.49.405}%
  \BibitemOpen
  \bibfield  {author} {\bibinfo {author} {\bibfnamefont {D.~J.}\ \bibnamefont
  {Thouless}}, \bibinfo {author} {\bibfnamefont {M.}~\bibnamefont {Kohmoto}},
  \bibinfo {author} {\bibfnamefont {M.~P.}\ \bibnamefont {Nightingale}},\ and\
  \bibinfo {author} {\bibfnamefont {M.}~\bibnamefont {den Nijs}},\ }\href
  {https://doi.org/10.1103/PhysRevLett.49.405} {\bibfield  {journal} {\bibinfo
  {journal} {Phys. Rev. Lett.}\ }\textbf {\bibinfo {volume} {49}},\ \bibinfo
  {pages} {405} (\bibinfo {year} {1982})}\BibitemShut {NoStop}%
\bibitem [{\citenamefont {Haldane}(1988)}]{PhysRevLett.61.2015}%
  \BibitemOpen
  \bibfield  {author} {\bibinfo {author} {\bibfnamefont {F.~D.~M.}\
  \bibnamefont {Haldane}},\ }\href
  {https://doi.org/10.1103/PhysRevLett.61.2015} {\bibfield  {journal} {\bibinfo
   {journal} {Phys. Rev. Lett.}\ }\textbf {\bibinfo {volume} {61}},\ \bibinfo
  {pages} {2015} (\bibinfo {year} {1988})}\BibitemShut {NoStop}%
\bibitem [{\citenamefont {Kane}\ and\ \citenamefont
  {Mele}(2005{\natexlab{a}})}]{PhysRevLett.95.146802}%
  \BibitemOpen
  \bibfield  {author} {\bibinfo {author} {\bibfnamefont {C.~L.}\ \bibnamefont
  {Kane}}\ and\ \bibinfo {author} {\bibfnamefont {E.~J.}\ \bibnamefont
  {Mele}},\ }\href {https://doi.org/10.1103/PhysRevLett.95.146802} {\bibfield
  {journal} {\bibinfo  {journal} {Phys. Rev. Lett.}\ }\textbf {\bibinfo
  {volume} {95}},\ \bibinfo {pages} {146802} (\bibinfo {year}
  {2005}{\natexlab{a}})}\BibitemShut {NoStop}%
\bibitem [{\citenamefont {Kane}\ and\ \citenamefont
  {Mele}(2005{\natexlab{b}})}]{PhysRevLett.95.226801}%
  \BibitemOpen
  \bibfield  {author} {\bibinfo {author} {\bibfnamefont {C.~L.}\ \bibnamefont
  {Kane}}\ and\ \bibinfo {author} {\bibfnamefont {E.~J.}\ \bibnamefont
  {Mele}},\ }\href {https://doi.org/10.1103/PhysRevLett.95.226801} {\bibfield
  {journal} {\bibinfo  {journal} {Phys. Rev. Lett.}\ }\textbf {\bibinfo
  {volume} {95}},\ \bibinfo {pages} {226801} (\bibinfo {year}
  {2005}{\natexlab{b}})}\BibitemShut {NoStop}%
\bibitem [{\citenamefont {Bernevig}\ \emph {et~al.}(2006)\citenamefont
  {Bernevig}, \citenamefont {Hughes},\ and\ \citenamefont
  {Zhang}}]{Bernevig1757}%
  \BibitemOpen
  \bibfield  {author} {\bibinfo {author} {\bibfnamefont {B.~A.}\ \bibnamefont
  {Bernevig}}, \bibinfo {author} {\bibfnamefont {T.~L.}\ \bibnamefont
  {Hughes}},\ and\ \bibinfo {author} {\bibfnamefont {S.-C.}\ \bibnamefont
  {Zhang}},\ }\href {https://doi.org/10.1126/science.1133734} {\bibfield
  {journal} {\bibinfo  {journal} {Science}\ }\textbf {\bibinfo {volume}
  {314}},\ \bibinfo {pages} {1757} (\bibinfo {year} {2006})}\BibitemShut
  {NoStop}%
\bibitem [{\citenamefont {Hasan}\ and\ \citenamefont
  {Kane}(2010)}]{RevModPhys.82.3045}%
  \BibitemOpen
  \bibfield  {author} {\bibinfo {author} {\bibfnamefont {M.~Z.}\ \bibnamefont
  {Hasan}}\ and\ \bibinfo {author} {\bibfnamefont {C.~L.}\ \bibnamefont
  {Kane}},\ }\href {https://doi.org/10.1103/RevModPhys.82.3045} {\bibfield
  {journal} {\bibinfo  {journal} {Rev. Mod. Phys.}\ }\textbf {\bibinfo {volume}
  {82}},\ \bibinfo {pages} {3045} (\bibinfo {year} {2010})}\BibitemShut
  {NoStop}%
\bibitem [{\citenamefont {Qi}\ and\ \citenamefont
  {Zhang}(2011)}]{RevModPhys.83.1057}%
  \BibitemOpen
  \bibfield  {author} {\bibinfo {author} {\bibfnamefont {X.-L.}\ \bibnamefont
  {Qi}}\ and\ \bibinfo {author} {\bibfnamefont {S.-C.}\ \bibnamefont {Zhang}},\
  }\href {https://doi.org/10.1103/RevModPhys.83.1057} {\bibfield  {journal}
  {\bibinfo  {journal} {Rev. Mod. Phys.}\ }\textbf {\bibinfo {volume} {83}},\
  \bibinfo {pages} {1057} (\bibinfo {year} {2011})}\BibitemShut {NoStop}%
\bibitem [{\citenamefont {Halperin}(1982)}]{PhysRevB.25.2185}%
  \BibitemOpen
  \bibfield  {author} {\bibinfo {author} {\bibfnamefont {B.~I.}\ \bibnamefont
  {Halperin}},\ }\href {https://doi.org/10.1103/PhysRevB.25.2185} {\bibfield
  {journal} {\bibinfo  {journal} {Phys. Rev. B}\ }\textbf {\bibinfo {volume}
  {25}},\ \bibinfo {pages} {2185} (\bibinfo {year} {1982})}\BibitemShut
  {NoStop}%
\bibitem [{\citenamefont {Hatsugai}(1993{\natexlab{a}})}]{PhysRevLett.71.3697}%
  \BibitemOpen
  \bibfield  {author} {\bibinfo {author} {\bibfnamefont {Y.}~\bibnamefont
  {Hatsugai}},\ }\href {https://doi.org/10.1103/PhysRevLett.71.3697} {\bibfield
   {journal} {\bibinfo  {journal} {Phys. Rev. Lett.}\ }\textbf {\bibinfo
  {volume} {71}},\ \bibinfo {pages} {3697} (\bibinfo {year}
  {1993}{\natexlab{a}})}\BibitemShut {NoStop}%
\bibitem [{\citenamefont {Hatsugai}(1993{\natexlab{b}})}]{PhysRevB.48.11851}%
  \BibitemOpen
  \bibfield  {author} {\bibinfo {author} {\bibfnamefont {Y.}~\bibnamefont
  {Hatsugai}},\ }\href {https://doi.org/10.1103/PhysRevB.48.11851} {\bibfield
  {journal} {\bibinfo  {journal} {Phys. Rev. B}\ }\textbf {\bibinfo {volume}
  {48}},\ \bibinfo {pages} {11851} (\bibinfo {year}
  {1993}{\natexlab{b}})}\BibitemShut {NoStop}%
\bibitem [{\citenamefont {Zak}(1989)}]{PhysRevLett.62.2747}%
  \BibitemOpen
  \bibfield  {author} {\bibinfo {author} {\bibfnamefont {J.}~\bibnamefont
  {Zak}},\ }\href {https://doi.org/10.1103/PhysRevLett.62.2747} {\bibfield
  {journal} {\bibinfo  {journal} {Phys. Rev. Lett.}\ }\textbf {\bibinfo
  {volume} {62}},\ \bibinfo {pages} {2747} (\bibinfo {year}
  {1989})}\BibitemShut {NoStop}%
\bibitem [{\citenamefont {King-Smith}\ and\ \citenamefont
  {Vanderbilt}(1993)}]{PhysRevB.47.1651}%
  \BibitemOpen
  \bibfield  {author} {\bibinfo {author} {\bibfnamefont {R.~D.}\ \bibnamefont
  {King-Smith}}\ and\ \bibinfo {author} {\bibfnamefont {D.}~\bibnamefont
  {Vanderbilt}},\ }\href {https://doi.org/10.1103/PhysRevB.47.1651} {\bibfield
  {journal} {\bibinfo  {journal} {Phys. Rev. B}\ }\textbf {\bibinfo {volume}
  {47}},\ \bibinfo {pages} {1651} (\bibinfo {year} {1993})}\BibitemShut
  {NoStop}%
\bibitem [{\citenamefont {Vanderbilt}\ and\ \citenamefont
  {King-Smith}(1993)}]{PhysRevB.48.4442}%
  \BibitemOpen
  \bibfield  {author} {\bibinfo {author} {\bibfnamefont {D.}~\bibnamefont
  {Vanderbilt}}\ and\ \bibinfo {author} {\bibfnamefont {R.~D.}\ \bibnamefont
  {King-Smith}},\ }\href {https://doi.org/10.1103/PhysRevB.48.4442} {\bibfield
  {journal} {\bibinfo  {journal} {Phys. Rev. B}\ }\textbf {\bibinfo {volume}
  {48}},\ \bibinfo {pages} {4442} (\bibinfo {year} {1993})}\BibitemShut
  {NoStop}%
\bibitem [{\citenamefont {Resta}(1994)}]{RevModPhys.66.899}%
  \BibitemOpen
  \bibfield  {author} {\bibinfo {author} {\bibfnamefont {R.}~\bibnamefont
  {Resta}},\ }\href {https://doi.org/10.1103/RevModPhys.66.899} {\bibfield
  {journal} {\bibinfo  {journal} {Rev. Mod. Phys.}\ }\textbf {\bibinfo {volume}
  {66}},\ \bibinfo {pages} {899} (\bibinfo {year} {1994})}\BibitemShut
  {NoStop}%
\bibitem [{\citenamefont {Ryu}\ and\ \citenamefont {Hatsugai}(2002)}]{Ryu2002}%
  \BibitemOpen
  \bibfield  {author} {\bibinfo {author} {\bibfnamefont {S.}~\bibnamefont
  {Ryu}}\ and\ \bibinfo {author} {\bibfnamefont {Y.}~\bibnamefont {Hatsugai}},\
  }\href {https://doi.org/10.1103/PhysRevLett.89.077002} {\bibfield  {journal}
  {\bibinfo  {journal} {Phys. Rev. Lett.}\ }\textbf {\bibinfo {volume} {89}},\
  \bibinfo {pages} {077002} (\bibinfo {year} {2002})}\BibitemShut {NoStop}%
\bibitem [{\citenamefont {Souza}\ \emph {et~al.}(2004)\citenamefont {Souza},
  \citenamefont {\'I\~niguez},\ and\ \citenamefont
  {Vanderbilt}}]{PhysRevB.69.085106}%
  \BibitemOpen
  \bibfield  {author} {\bibinfo {author} {\bibfnamefont {I.}~\bibnamefont
  {Souza}}, \bibinfo {author} {\bibfnamefont {J.}~\bibnamefont {\'I\~niguez}},\
  and\ \bibinfo {author} {\bibfnamefont {D.}~\bibnamefont {Vanderbilt}},\
  }\href {https://doi.org/10.1103/PhysRevB.69.085106} {\bibfield  {journal}
  {\bibinfo  {journal} {Phys. Rev. B}\ }\textbf {\bibinfo {volume} {69}},\
  \bibinfo {pages} {085106} (\bibinfo {year} {2004})}\BibitemShut {NoStop}%
\bibitem [{\citenamefont {Qi}\ \emph {et~al.}(2008)\citenamefont {Qi},
  \citenamefont {Hughes},\ and\ \citenamefont {Zhang}}]{PhysRevB.78.195424}%
  \BibitemOpen
  \bibfield  {author} {\bibinfo {author} {\bibfnamefont {X.-L.}\ \bibnamefont
  {Qi}}, \bibinfo {author} {\bibfnamefont {T.~L.}\ \bibnamefont {Hughes}},\
  and\ \bibinfo {author} {\bibfnamefont {S.-C.}\ \bibnamefont {Zhang}},\ }\href
  {https://doi.org/10.1103/PhysRevB.78.195424} {\bibfield  {journal} {\bibinfo
  {journal} {Phys. Rev. B}\ }\textbf {\bibinfo {volume} {78}},\ \bibinfo
  {pages} {195424} (\bibinfo {year} {2008})}\BibitemShut {NoStop}%
\bibitem [{\citenamefont {Hatsugai}(2009)}]{HATSUGAI20091061}%
  \BibitemOpen
  \bibfield  {author} {\bibinfo {author} {\bibfnamefont {Y.}~\bibnamefont
  {Hatsugai}},\ }\href
  {https://doi.org/https://doi.org/10.1016/j.ssc.2009.02.055} {\bibfield
  {journal} {\bibinfo  {journal} {Solid State Communications}\ }\textbf
  {\bibinfo {volume} {149}},\ \bibinfo {pages} {1061 } (\bibinfo {year}
  {2009})},\ \bibinfo {note} {recent Progress in Graphene Studies}\BibitemShut
  {NoStop}%
\bibitem [{\citenamefont {Fang}\ \emph {et~al.}(2012)\citenamefont {Fang},
  \citenamefont {Gilbert},\ and\ \citenamefont
  {Bernevig}}]{PhysRevB.86.115112}%
  \BibitemOpen
  \bibfield  {author} {\bibinfo {author} {\bibfnamefont {C.}~\bibnamefont
  {Fang}}, \bibinfo {author} {\bibfnamefont {M.~J.}\ \bibnamefont {Gilbert}},\
  and\ \bibinfo {author} {\bibfnamefont {B.~A.}\ \bibnamefont {Bernevig}},\
  }\href {https://doi.org/10.1103/PhysRevB.86.115112} {\bibfield  {journal}
  {\bibinfo  {journal} {Phys. Rev. B}\ }\textbf {\bibinfo {volume} {86}},\
  \bibinfo {pages} {115112} (\bibinfo {year} {2012})}\BibitemShut {NoStop}%
\bibitem [{\citenamefont {Watanabe}\ and\ \citenamefont
  {Oshikawa}(2018)}]{PhysRevX.8.021065}%
  \BibitemOpen
  \bibfield  {author} {\bibinfo {author} {\bibfnamefont {H.}~\bibnamefont
  {Watanabe}}\ and\ \bibinfo {author} {\bibfnamefont {M.}~\bibnamefont
  {Oshikawa}},\ }\href {https://doi.org/10.1103/PhysRevX.8.021065} {\bibfield
  {journal} {\bibinfo  {journal} {Phys. Rev. X}\ }\textbf {\bibinfo {volume}
  {8}},\ \bibinfo {pages} {021065} (\bibinfo {year} {2018})}\BibitemShut
  {NoStop}%
\bibitem [{\citenamefont {Hatsugai}\ and\ \citenamefont
  {Maruyama}(2011)}]{0295-5075-95-2-20003}%
  \BibitemOpen
  \bibfield  {author} {\bibinfo {author} {\bibfnamefont {Y.}~\bibnamefont
  {Hatsugai}}\ and\ \bibinfo {author} {\bibfnamefont {I.}~\bibnamefont
  {Maruyama}},\ }\href {http://stacks.iop.org/0295-5075/95/i=2/a=20003}
  {\bibfield  {journal} {\bibinfo  {journal} {Europhys. Lett.}\ }\textbf
  {\bibinfo {volume} {95}},\ \bibinfo {pages} {20003} (\bibinfo {year}
  {2011})}\BibitemShut {NoStop}%
\bibitem [{\citenamefont {Kariyado}\ and\ \citenamefont
  {Hatsugai}(2014)}]{PhysRevB.90.085132}%
  \BibitemOpen
  \bibfield  {author} {\bibinfo {author} {\bibfnamefont {T.}~\bibnamefont
  {Kariyado}}\ and\ \bibinfo {author} {\bibfnamefont {Y.}~\bibnamefont
  {Hatsugai}},\ }\href {https://doi.org/10.1103/PhysRevB.90.085132} {\bibfield
  {journal} {\bibinfo  {journal} {Phys. Rev. B}\ }\textbf {\bibinfo {volume}
  {90}},\ \bibinfo {pages} {085132} (\bibinfo {year} {2014})}\BibitemShut
  {NoStop}%
\bibitem [{\citenamefont {Benalcazar}\ \emph
  {et~al.}(2017{\natexlab{a}})\citenamefont {Benalcazar}, \citenamefont
  {Bernevig},\ and\ \citenamefont {Hughes}}]{PhysRevB.96.245115}%
  \BibitemOpen
  \bibfield  {author} {\bibinfo {author} {\bibfnamefont {W.~A.}\ \bibnamefont
  {Benalcazar}}, \bibinfo {author} {\bibfnamefont {B.~A.}\ \bibnamefont
  {Bernevig}},\ and\ \bibinfo {author} {\bibfnamefont {T.~L.}\ \bibnamefont
  {Hughes}},\ }\href {https://doi.org/10.1103/PhysRevB.96.245115} {\bibfield
  {journal} {\bibinfo  {journal} {Phys. Rev. B}\ }\textbf {\bibinfo {volume}
  {96}},\ \bibinfo {pages} {245115} (\bibinfo {year}
  {2017}{\natexlab{a}})}\BibitemShut {NoStop}%
\bibitem [{\citenamefont {Benalcazar}\ \emph
  {et~al.}(2017{\natexlab{b}})\citenamefont {Benalcazar}, \citenamefont
  {Bernevig},\ and\ \citenamefont {Hughes}}]{Benalcazar61}%
  \BibitemOpen
  \bibfield  {author} {\bibinfo {author} {\bibfnamefont {W.~A.}\ \bibnamefont
  {Benalcazar}}, \bibinfo {author} {\bibfnamefont {B.~A.}\ \bibnamefont
  {Bernevig}},\ and\ \bibinfo {author} {\bibfnamefont {T.~L.}\ \bibnamefont
  {Hughes}},\ }\href {https://doi.org/10.1126/science.aah6442} {\bibfield
  {journal} {\bibinfo  {journal} {Science}\ }\textbf {\bibinfo {volume}
  {357}},\ \bibinfo {pages} {61} (\bibinfo {year}
  {2017}{\natexlab{b}})}\BibitemShut {NoStop}%
\bibitem [{\citenamefont {Ono}\ \emph {et~al.}(2019)\citenamefont {Ono},
  \citenamefont {Trifunovic},\ and\ \citenamefont
  {Watanabe}}]{PhysRevB.100.245133}%
  \BibitemOpen
  \bibfield  {author} {\bibinfo {author} {\bibfnamefont {S.}~\bibnamefont
  {Ono}}, \bibinfo {author} {\bibfnamefont {L.}~\bibnamefont {Trifunovic}},\
  and\ \bibinfo {author} {\bibfnamefont {H.}~\bibnamefont {Watanabe}},\ }\href
  {https://doi.org/10.1103/PhysRevB.100.245133} {\bibfield  {journal} {\bibinfo
   {journal} {Phys. Rev. B}\ }\textbf {\bibinfo {volume} {100}},\ \bibinfo
  {pages} {245133} (\bibinfo {year} {2019})}\BibitemShut {NoStop}%
\bibitem [{\citenamefont {Kang}\ \emph {et~al.}(2019)\citenamefont {Kang},
  \citenamefont {Shiozaki},\ and\ \citenamefont {Cho}}]{PhysRevB.100.245134}%
  \BibitemOpen
  \bibfield  {author} {\bibinfo {author} {\bibfnamefont {B.}~\bibnamefont
  {Kang}}, \bibinfo {author} {\bibfnamefont {K.}~\bibnamefont {Shiozaki}},\
  and\ \bibinfo {author} {\bibfnamefont {G.~Y.}\ \bibnamefont {Cho}},\ }\href
  {https://doi.org/10.1103/PhysRevB.100.245134} {\bibfield  {journal} {\bibinfo
   {journal} {Phys. Rev. B}\ }\textbf {\bibinfo {volume} {100}},\ \bibinfo
  {pages} {245134} (\bibinfo {year} {2019})}\BibitemShut {NoStop}%
\bibitem [{\citenamefont {Wheeler}\ \emph {et~al.}(2019)\citenamefont
  {Wheeler}, \citenamefont {Wagner},\ and\ \citenamefont
  {Hughes}}]{PhysRevB.100.245135}%
  \BibitemOpen
  \bibfield  {author} {\bibinfo {author} {\bibfnamefont {W.~A.}\ \bibnamefont
  {Wheeler}}, \bibinfo {author} {\bibfnamefont {L.~K.}\ \bibnamefont
  {Wagner}},\ and\ \bibinfo {author} {\bibfnamefont {T.~L.}\ \bibnamefont
  {Hughes}},\ }\href {https://doi.org/10.1103/PhysRevB.100.245135} {\bibfield
  {journal} {\bibinfo  {journal} {Phys. Rev. B}\ }\textbf {\bibinfo {volume}
  {100}},\ \bibinfo {pages} {245135} (\bibinfo {year} {2019})}\BibitemShut
  {NoStop}%
\bibitem [{\citenamefont {Watanabe}\ and\ \citenamefont
  {Ono}(2020)}]{Watanabe2020}%
  \BibitemOpen
  \bibfield  {author} {\bibinfo {author} {\bibfnamefont {H.}~\bibnamefont
  {Watanabe}}\ and\ \bibinfo {author} {\bibfnamefont {S.}~\bibnamefont {Ono}},\
  }\href {https://doi.org/10.1103/PhysRevB.102.165120} {\bibfield  {journal}
  {\bibinfo  {journal} {Phys. Rev. B}\ }\textbf {\bibinfo {volume} {102}},\
  \bibinfo {pages} {165120} (\bibinfo {year} {2020})}\BibitemShut {NoStop}%
\bibitem [{\citenamefont {Hashimoto}\ \emph {et~al.}(2017)\citenamefont
  {Hashimoto}, \citenamefont {Wu},\ and\ \citenamefont
  {Kimura}}]{PhysRevB.95.165443}%
  \BibitemOpen
  \bibfield  {author} {\bibinfo {author} {\bibfnamefont {K.}~\bibnamefont
  {Hashimoto}}, \bibinfo {author} {\bibfnamefont {X.}~\bibnamefont {Wu}},\ and\
  \bibinfo {author} {\bibfnamefont {T.}~\bibnamefont {Kimura}},\ }\href
  {https://doi.org/10.1103/PhysRevB.95.165443} {\bibfield  {journal} {\bibinfo
  {journal} {Phys. Rev. B}\ }\textbf {\bibinfo {volume} {95}},\ \bibinfo
  {pages} {165443} (\bibinfo {year} {2017})}\BibitemShut {NoStop}%
\bibitem [{\citenamefont {Hayashi}(2018)}]{Hayashi2018}%
  \BibitemOpen
  \bibfield  {author} {\bibinfo {author} {\bibfnamefont {S.}~\bibnamefont
  {Hayashi}},\ }\href {https://doi.org/10.1007/s00220-018-3229-2} {\bibfield
  {journal} {\bibinfo  {journal} {Communications in Mathematical Physics}\
  }\textbf {\bibinfo {volume} {364}},\ \bibinfo {pages} {343} (\bibinfo {year}
  {2018})}\BibitemShut {NoStop}%
\bibitem [{\citenamefont {Xu}\ \emph {et~al.}()\citenamefont {Xu},
  \citenamefont {Xue},\ and\ \citenamefont {Wan}}]{Xu2017}%
  \BibitemOpen
  \bibfield  {author} {\bibinfo {author} {\bibfnamefont {Y.}~\bibnamefont
  {Xu}}, \bibinfo {author} {\bibfnamefont {R.}~\bibnamefont {Xue}},\ and\
  \bibinfo {author} {\bibfnamefont {S.}~\bibnamefont {Wan}},\ }\href@noop {} {\
  }\Eprint {https://arxiv.org/abs/1711.09202} {arXiv:1711.09202
  [cond-mat.str-el]} \BibitemShut {NoStop}%
\bibitem [{\citenamefont {Schindler}\ \emph
  {et~al.}(2018{\natexlab{a}})\citenamefont {Schindler}, \citenamefont {Cook},
  \citenamefont {Vergniory}, \citenamefont {Wang}, \citenamefont {Parkin},
  \citenamefont {Bernevig},\ and\ \citenamefont {Neupert}}]{Schindlereaat0346}%
  \BibitemOpen
  \bibfield  {author} {\bibinfo {author} {\bibfnamefont {F.}~\bibnamefont
  {Schindler}}, \bibinfo {author} {\bibfnamefont {A.~M.}\ \bibnamefont {Cook}},
  \bibinfo {author} {\bibfnamefont {M.~G.}\ \bibnamefont {Vergniory}}, \bibinfo
  {author} {\bibfnamefont {Z.}~\bibnamefont {Wang}}, \bibinfo {author}
  {\bibfnamefont {S.~S.~P.}\ \bibnamefont {Parkin}}, \bibinfo {author}
  {\bibfnamefont {B.~A.}\ \bibnamefont {Bernevig}},\ and\ \bibinfo {author}
  {\bibfnamefont {T.}~\bibnamefont {Neupert}},\ }\bibfield  {journal} {\bibinfo
   {journal} {Science Advances}\ }\textbf {\bibinfo {volume} {4}},\ \href
  {https://doi.org/10.1126/sciadv.aat0346} {10.1126/sciadv.aat0346} (\bibinfo
  {year} {2018}{\natexlab{a}})\BibitemShut {NoStop}%
\bibitem [{\citenamefont {Ezawa}(2018)}]{PhysRevLett.120.026801}%
  \BibitemOpen
  \bibfield  {author} {\bibinfo {author} {\bibfnamefont {M.}~\bibnamefont
  {Ezawa}},\ }\href {https://doi.org/10.1103/PhysRevLett.120.026801} {\bibfield
   {journal} {\bibinfo  {journal} {Phys. Rev. Lett.}\ }\textbf {\bibinfo
  {volume} {120}},\ \bibinfo {pages} {026801} (\bibinfo {year}
  {2018})}\BibitemShut {NoStop}%
\bibitem [{\citenamefont {Khalaf}(2018)}]{PhysRevB.97.205136}%
  \BibitemOpen
  \bibfield  {author} {\bibinfo {author} {\bibfnamefont {E.}~\bibnamefont
  {Khalaf}},\ }\href {https://doi.org/10.1103/PhysRevB.97.205136} {\bibfield
  {journal} {\bibinfo  {journal} {Phys. Rev. B}\ }\textbf {\bibinfo {volume}
  {97}},\ \bibinfo {pages} {205136} (\bibinfo {year} {2018})}\BibitemShut
  {NoStop}%
\bibitem [{\citenamefont {Kunst}\ \emph {et~al.}(2018)\citenamefont {Kunst},
  \citenamefont {van Miert},\ and\ \citenamefont
  {Bergholtz}}]{PhysRevB.97.241405}%
  \BibitemOpen
  \bibfield  {author} {\bibinfo {author} {\bibfnamefont {F.~K.}\ \bibnamefont
  {Kunst}}, \bibinfo {author} {\bibfnamefont {G.}~\bibnamefont {van Miert}},\
  and\ \bibinfo {author} {\bibfnamefont {E.~J.}\ \bibnamefont {Bergholtz}},\
  }\href {https://doi.org/10.1103/PhysRevB.97.241405} {\bibfield  {journal}
  {\bibinfo  {journal} {Phys. Rev. B}\ }\textbf {\bibinfo {volume} {97}},\
  \bibinfo {pages} {241405} (\bibinfo {year} {2018})}\BibitemShut {NoStop}%
\bibitem [{\citenamefont {Fukui}\ and\ \citenamefont
  {Hatsugai}(2018)}]{PhysRevB.98.035147}%
  \BibitemOpen
  \bibfield  {author} {\bibinfo {author} {\bibfnamefont {T.}~\bibnamefont
  {Fukui}}\ and\ \bibinfo {author} {\bibfnamefont {Y.}~\bibnamefont
  {Hatsugai}},\ }\href {https://doi.org/10.1103/PhysRevB.98.035147} {\bibfield
  {journal} {\bibinfo  {journal} {Phys. Rev. B}\ }\textbf {\bibinfo {volume}
  {98}},\ \bibinfo {pages} {035147} (\bibinfo {year} {2018})}\BibitemShut
  {NoStop}%
\bibitem [{\citenamefont {You}\ \emph {et~al.}(2018)\citenamefont {You},
  \citenamefont {Devakul}, \citenamefont {Burnell},\ and\ \citenamefont
  {Neupert}}]{PhysRevB.98.235102}%
  \BibitemOpen
  \bibfield  {author} {\bibinfo {author} {\bibfnamefont {Y.}~\bibnamefont
  {You}}, \bibinfo {author} {\bibfnamefont {T.}~\bibnamefont {Devakul}},
  \bibinfo {author} {\bibfnamefont {F.~J.}\ \bibnamefont {Burnell}},\ and\
  \bibinfo {author} {\bibfnamefont {T.}~\bibnamefont {Neupert}},\ }\href
  {https://doi.org/10.1103/PhysRevB.98.235102} {\bibfield  {journal} {\bibinfo
  {journal} {Phys. Rev. B}\ }\textbf {\bibinfo {volume} {98}},\ \bibinfo
  {pages} {235102} (\bibinfo {year} {2018})}\BibitemShut {NoStop}%
\bibitem [{\citenamefont {C\ifmmode \u{a}\else \u{a}\fi{}lug\ifmmode~\u{a}\else
  \u{a}\fi{}ru}\ \emph {et~al.}(2019)\citenamefont {C\ifmmode \u{a}\else
  \u{a}\fi{}lug\ifmmode~\u{a}\else \u{a}\fi{}ru}, \citenamefont {Juri\ifmmode
  \check{c}\else \v{c}\fi{}i\ifmmode~\acute{c}\else \'{c}\fi{}},\ and\
  \citenamefont {Roy}}]{PhysRevB.99.041301}%
  \BibitemOpen
  \bibfield  {author} {\bibinfo {author} {\bibfnamefont {D.}~\bibnamefont
  {C\ifmmode \u{a}\else \u{a}\fi{}lug\ifmmode~\u{a}\else \u{a}\fi{}ru}},
  \bibinfo {author} {\bibfnamefont {V.}~\bibnamefont {Juri\ifmmode
  \check{c}\else \v{c}\fi{}i\ifmmode~\acute{c}\else \'{c}\fi{}}},\ and\
  \bibinfo {author} {\bibfnamefont {B.}~\bibnamefont {Roy}},\ }\href
  {https://doi.org/10.1103/PhysRevB.99.041301} {\bibfield  {journal} {\bibinfo
  {journal} {Phys. Rev. B}\ }\textbf {\bibinfo {volume} {99}},\ \bibinfo
  {pages} {041301} (\bibinfo {year} {2019})}\BibitemShut {NoStop}%
\bibitem [{\citenamefont {Okuma}\ \emph {et~al.}(2019)\citenamefont {Okuma},
  \citenamefont {Sato},\ and\ \citenamefont {Shiozaki}}]{PhysRevB.99.085127}%
  \BibitemOpen
  \bibfield  {author} {\bibinfo {author} {\bibfnamefont {N.}~\bibnamefont
  {Okuma}}, \bibinfo {author} {\bibfnamefont {M.}~\bibnamefont {Sato}},\ and\
  \bibinfo {author} {\bibfnamefont {K.}~\bibnamefont {Shiozaki}},\ }\href
  {https://doi.org/10.1103/PhysRevB.99.085127} {\bibfield  {journal} {\bibinfo
  {journal} {Phys. Rev. B}\ }\textbf {\bibinfo {volume} {99}},\ \bibinfo
  {pages} {085127} (\bibinfo {year} {2019})}\BibitemShut {NoStop}%
\bibitem [{\citenamefont {Araki}\ \emph {et~al.}(2019)\citenamefont {Araki},
  \citenamefont {Mizoguchi},\ and\ \citenamefont
  {Hatsugai}}]{PhysRevB.99.085406}%
  \BibitemOpen
  \bibfield  {author} {\bibinfo {author} {\bibfnamefont {H.}~\bibnamefont
  {Araki}}, \bibinfo {author} {\bibfnamefont {T.}~\bibnamefont {Mizoguchi}},\
  and\ \bibinfo {author} {\bibfnamefont {Y.}~\bibnamefont {Hatsugai}},\ }\href
  {https://doi.org/10.1103/PhysRevB.99.085406} {\bibfield  {journal} {\bibinfo
  {journal} {Phys. Rev. B}\ }\textbf {\bibinfo {volume} {99}},\ \bibinfo
  {pages} {085406} (\bibinfo {year} {2019})}\BibitemShut {NoStop}%
\bibitem [{\citenamefont {Dubinkin}\ and\ \citenamefont
  {Hughes}(2019)}]{PhysRevB.99.235132}%
  \BibitemOpen
  \bibfield  {author} {\bibinfo {author} {\bibfnamefont {O.}~\bibnamefont
  {Dubinkin}}\ and\ \bibinfo {author} {\bibfnamefont {T.~L.}\ \bibnamefont
  {Hughes}},\ }\href {https://doi.org/10.1103/PhysRevB.99.235132} {\bibfield
  {journal} {\bibinfo  {journal} {Phys. Rev. B}\ }\textbf {\bibinfo {volume}
  {99}},\ \bibinfo {pages} {235132} (\bibinfo {year} {2019})}\BibitemShut
  {NoStop}%
\bibitem [{\citenamefont {Benalcazar}\ \emph {et~al.}(2019)\citenamefont
  {Benalcazar}, \citenamefont {Li},\ and\ \citenamefont
  {Hughes}}]{PhysRevB.99.245151}%
  \BibitemOpen
  \bibfield  {author} {\bibinfo {author} {\bibfnamefont {W.~A.}\ \bibnamefont
  {Benalcazar}}, \bibinfo {author} {\bibfnamefont {T.}~\bibnamefont {Li}},\
  and\ \bibinfo {author} {\bibfnamefont {T.~L.}\ \bibnamefont {Hughes}},\
  }\href {https://doi.org/10.1103/PhysRevB.99.245151} {\bibfield  {journal}
  {\bibinfo  {journal} {Phys. Rev. B}\ }\textbf {\bibinfo {volume} {99}},\
  \bibinfo {pages} {245151} (\bibinfo {year} {2019})}\BibitemShut {NoStop}%
\bibitem [{\citenamefont {Kudo}\ \emph {et~al.}(2019)\citenamefont {Kudo},
  \citenamefont {Yoshida},\ and\ \citenamefont
  {Hatsugai}}]{PhysRevLett.123.196402}%
  \BibitemOpen
  \bibfield  {author} {\bibinfo {author} {\bibfnamefont {K.}~\bibnamefont
  {Kudo}}, \bibinfo {author} {\bibfnamefont {T.}~\bibnamefont {Yoshida}},\ and\
  \bibinfo {author} {\bibfnamefont {Y.}~\bibnamefont {Hatsugai}},\ }\href
  {https://doi.org/10.1103/PhysRevLett.123.196402} {\bibfield  {journal}
  {\bibinfo  {journal} {Phys. Rev. Lett.}\ }\textbf {\bibinfo {volume} {123}},\
  \bibinfo {pages} {196402} (\bibinfo {year} {2019})}\BibitemShut {NoStop}%
\bibitem [{\citenamefont {Okugawa}\ \emph {et~al.}(2019)\citenamefont
  {Okugawa}, \citenamefont {Hayashi},\ and\ \citenamefont
  {Nakanishi}}]{PhysRevB.100.235302}%
  \BibitemOpen
  \bibfield  {author} {\bibinfo {author} {\bibfnamefont {R.}~\bibnamefont
  {Okugawa}}, \bibinfo {author} {\bibfnamefont {S.}~\bibnamefont {Hayashi}},\
  and\ \bibinfo {author} {\bibfnamefont {T.}~\bibnamefont {Nakanishi}},\ }\href
  {https://doi.org/10.1103/PhysRevB.100.235302} {\bibfield  {journal} {\bibinfo
   {journal} {Phys. Rev. B}\ }\textbf {\bibinfo {volume} {100}},\ \bibinfo
  {pages} {235302} (\bibinfo {year} {2019})}\BibitemShut {NoStop}%
\bibitem [{\citenamefont {Araki}\ \emph {et~al.}(2020)\citenamefont {Araki},
  \citenamefont {Mizoguchi},\ and\ \citenamefont
  {Hatsugai}}]{PhysRevResearch.2.012009}%
  \BibitemOpen
  \bibfield  {author} {\bibinfo {author} {\bibfnamefont {H.}~\bibnamefont
  {Araki}}, \bibinfo {author} {\bibfnamefont {T.}~\bibnamefont {Mizoguchi}},\
  and\ \bibinfo {author} {\bibfnamefont {Y.}~\bibnamefont {Hatsugai}},\ }\href
  {https://doi.org/10.1103/PhysRevResearch.2.012009} {\bibfield  {journal}
  {\bibinfo  {journal} {Phys. Rev. Research}\ }\textbf {\bibinfo {volume}
  {2}},\ \bibinfo {pages} {012009} (\bibinfo {year} {2020})}\BibitemShut
  {NoStop}%
\bibitem [{\citenamefont {Rasmussen}\ and\ \citenamefont
  {Lu}(2020)}]{PhysRevB.101.085137}%
  \BibitemOpen
  \bibfield  {author} {\bibinfo {author} {\bibfnamefont {A.}~\bibnamefont
  {Rasmussen}}\ and\ \bibinfo {author} {\bibfnamefont {Y.-M.}\ \bibnamefont
  {Lu}},\ }\href {https://doi.org/10.1103/PhysRevB.101.085137} {\bibfield
  {journal} {\bibinfo  {journal} {Phys. Rev. B}\ }\textbf {\bibinfo {volume}
  {101}},\ \bibinfo {pages} {085137} (\bibinfo {year} {2020})}\BibitemShut
  {NoStop}%
\bibitem [{\citenamefont {Zhu}\ \emph {et~al.}(2020)\citenamefont {Zhu},
  \citenamefont {Loehr},\ and\ \citenamefont {Hughes}}]{PhysRevB.101.115140}%
  \BibitemOpen
  \bibfield  {author} {\bibinfo {author} {\bibfnamefont {P.}~\bibnamefont
  {Zhu}}, \bibinfo {author} {\bibfnamefont {K.}~\bibnamefont {Loehr}},\ and\
  \bibinfo {author} {\bibfnamefont {T.~L.}\ \bibnamefont {Hughes}},\ }\href
  {https://doi.org/10.1103/PhysRevB.101.115140} {\bibfield  {journal} {\bibinfo
   {journal} {Phys. Rev. B}\ }\textbf {\bibinfo {volume} {101}},\ \bibinfo
  {pages} {115140} (\bibinfo {year} {2020})}\BibitemShut {NoStop}%
\bibitem [{\citenamefont {Schindler}\ \emph
  {et~al.}(2018{\natexlab{b}})\citenamefont {Schindler}, \citenamefont {Wang},
  \citenamefont {Vergniory}, \citenamefont {Cook}, \citenamefont {Murani},
  \citenamefont {Sengupta}, \citenamefont {Kasumov}, \citenamefont {Deblock},
  \citenamefont {Jeon}, \citenamefont {Drozdov}, \citenamefont {Bouchiat},
  \citenamefont {Gu{\'e}ron}, \citenamefont {Yazdani}, \citenamefont
  {Bernevig},\ and\ \citenamefont {Neupert}}]{Schindler2018}%
  \BibitemOpen
  \bibfield  {author} {\bibinfo {author} {\bibfnamefont {F.}~\bibnamefont
  {Schindler}}, \bibinfo {author} {\bibfnamefont {Z.}~\bibnamefont {Wang}},
  \bibinfo {author} {\bibfnamefont {M.~G.}\ \bibnamefont {Vergniory}}, \bibinfo
  {author} {\bibfnamefont {A.~M.}\ \bibnamefont {Cook}}, \bibinfo {author}
  {\bibfnamefont {A.}~\bibnamefont {Murani}}, \bibinfo {author} {\bibfnamefont
  {S.}~\bibnamefont {Sengupta}}, \bibinfo {author} {\bibfnamefont {A.~Y.}\
  \bibnamefont {Kasumov}}, \bibinfo {author} {\bibfnamefont {R.}~\bibnamefont
  {Deblock}}, \bibinfo {author} {\bibfnamefont {S.}~\bibnamefont {Jeon}},
  \bibinfo {author} {\bibfnamefont {I.}~\bibnamefont {Drozdov}}, \bibinfo
  {author} {\bibfnamefont {H.}~\bibnamefont {Bouchiat}}, \bibinfo {author}
  {\bibfnamefont {S.}~\bibnamefont {Gu{\'e}ron}}, \bibinfo {author}
  {\bibfnamefont {A.}~\bibnamefont {Yazdani}}, \bibinfo {author} {\bibfnamefont
  {B.~A.}\ \bibnamefont {Bernevig}},\ and\ \bibinfo {author} {\bibfnamefont
  {T.}~\bibnamefont {Neupert}},\ }\href
  {https://doi.org/10.1038/s41567-018-0224-7} {\bibfield  {journal} {\bibinfo
  {journal} {Nature Physics}\ }\textbf {\bibinfo {volume} {14}},\ \bibinfo
  {pages} {918} (\bibinfo {year} {2018}{\natexlab{b}})}\BibitemShut {NoStop}%
\bibitem [{\citenamefont {Serra-Garcia}\ \emph {et~al.}(2018)\citenamefont
  {Serra-Garcia}, \citenamefont {Peri}, \citenamefont {S{\"u}sstrunk},
  \citenamefont {Bilal}, \citenamefont {Larsen}, \citenamefont {Villanueva},\
  and\ \citenamefont {Huber}}]{Garcia2018}%
  \BibitemOpen
  \bibfield  {author} {\bibinfo {author} {\bibfnamefont {M.}~\bibnamefont
  {Serra-Garcia}}, \bibinfo {author} {\bibfnamefont {V.}~\bibnamefont {Peri}},
  \bibinfo {author} {\bibfnamefont {R.}~\bibnamefont {S{\"u}sstrunk}}, \bibinfo
  {author} {\bibfnamefont {O.~R.}\ \bibnamefont {Bilal}}, \bibinfo {author}
  {\bibfnamefont {T.}~\bibnamefont {Larsen}}, \bibinfo {author} {\bibfnamefont
  {L.~G.}\ \bibnamefont {Villanueva}},\ and\ \bibinfo {author} {\bibfnamefont
  {S.~D.}\ \bibnamefont {Huber}},\ }\href {https://doi.org/10.1038/nature25156}
  {\bibfield  {journal} {\bibinfo  {journal} {Nature}\ }\textbf {\bibinfo
  {volume} {555}},\ \bibinfo {pages} {342 EP } (\bibinfo {year}
  {2018})}\BibitemShut {NoStop}%
\bibitem [{\citenamefont {Imhof}\ \emph {et~al.}(2018)\citenamefont {Imhof},
  \citenamefont {Berger}, \citenamefont {Bayer}, \citenamefont {Brehm},
  \citenamefont {Molenkamp}, \citenamefont {Kiessling}, \citenamefont
  {Schindler}, \citenamefont {Lee}, \citenamefont {Greiter}, \citenamefont
  {Neupert},\ and\ \citenamefont {Thomale}}]{Imhof2018}%
  \BibitemOpen
  \bibfield  {author} {\bibinfo {author} {\bibfnamefont {S.}~\bibnamefont
  {Imhof}}, \bibinfo {author} {\bibfnamefont {C.}~\bibnamefont {Berger}},
  \bibinfo {author} {\bibfnamefont {F.}~\bibnamefont {Bayer}}, \bibinfo
  {author} {\bibfnamefont {J.}~\bibnamefont {Brehm}}, \bibinfo {author}
  {\bibfnamefont {L.~W.}\ \bibnamefont {Molenkamp}}, \bibinfo {author}
  {\bibfnamefont {T.}~\bibnamefont {Kiessling}}, \bibinfo {author}
  {\bibfnamefont {F.}~\bibnamefont {Schindler}}, \bibinfo {author}
  {\bibfnamefont {C.~H.}\ \bibnamefont {Lee}}, \bibinfo {author} {\bibfnamefont
  {M.}~\bibnamefont {Greiter}}, \bibinfo {author} {\bibfnamefont
  {T.}~\bibnamefont {Neupert}},\ and\ \bibinfo {author} {\bibfnamefont
  {R.}~\bibnamefont {Thomale}},\ }\href
  {https://doi.org/10.1038/s41567-018-0246-1} {\bibfield  {journal} {\bibinfo
  {journal} {Nature Physics}\ }\textbf {\bibinfo {volume} {14}},\ \bibinfo
  {pages} {925} (\bibinfo {year} {2018})}\BibitemShut {NoStop}%
\bibitem [{\citenamefont {Xie}\ \emph {et~al.}(2018)\citenamefont {Xie},
  \citenamefont {Wang}, \citenamefont {Wang}, \citenamefont {Zhu},
  \citenamefont {Jiang}, \citenamefont {Lu},\ and\ \citenamefont
  {Chen}}]{PhysRevB.98.205147}%
  \BibitemOpen
  \bibfield  {author} {\bibinfo {author} {\bibfnamefont {B.-Y.}\ \bibnamefont
  {Xie}}, \bibinfo {author} {\bibfnamefont {H.-F.}\ \bibnamefont {Wang}},
  \bibinfo {author} {\bibfnamefont {H.-X.}\ \bibnamefont {Wang}}, \bibinfo
  {author} {\bibfnamefont {X.-Y.}\ \bibnamefont {Zhu}}, \bibinfo {author}
  {\bibfnamefont {J.-H.}\ \bibnamefont {Jiang}}, \bibinfo {author}
  {\bibfnamefont {M.-H.}\ \bibnamefont {Lu}},\ and\ \bibinfo {author}
  {\bibfnamefont {Y.-F.}\ \bibnamefont {Chen}},\ }\href
  {https://doi.org/10.1103/PhysRevB.98.205147} {\bibfield  {journal} {\bibinfo
  {journal} {Phys. Rev. B}\ }\textbf {\bibinfo {volume} {98}},\ \bibinfo
  {pages} {205147} (\bibinfo {year} {2018})}\BibitemShut {NoStop}%
\bibitem [{\citenamefont {Peterson}\ \emph {et~al.}(2018)\citenamefont
  {Peterson}, \citenamefont {Benalcazar}, \citenamefont {Hughes},\ and\
  \citenamefont {Bahl}}]{Peterson2018}%
  \BibitemOpen
  \bibfield  {author} {\bibinfo {author} {\bibfnamefont {C.~W.}\ \bibnamefont
  {Peterson}}, \bibinfo {author} {\bibfnamefont {W.~A.}\ \bibnamefont
  {Benalcazar}}, \bibinfo {author} {\bibfnamefont {T.~L.}\ \bibnamefont
  {Hughes}},\ and\ \bibinfo {author} {\bibfnamefont {G.}~\bibnamefont {Bahl}},\
  }\href {https://doi.org/10.1038/nature25777} {\bibfield  {journal} {\bibinfo
  {journal} {Nature}\ }\textbf {\bibinfo {volume} {555}},\ \bibinfo {pages}
  {346} (\bibinfo {year} {2018})}\BibitemShut {NoStop}%
\bibitem [{\citenamefont {Noh}\ \emph {et~al.}(2018)\citenamefont {Noh},
  \citenamefont {Benalcazar}, \citenamefont {Huang}, \citenamefont {Collins},
  \citenamefont {Chen}, \citenamefont {Hughes},\ and\ \citenamefont
  {Rechtsman}}]{Noh2018}%
  \BibitemOpen
  \bibfield  {author} {\bibinfo {author} {\bibfnamefont {J.}~\bibnamefont
  {Noh}}, \bibinfo {author} {\bibfnamefont {W.~A.}\ \bibnamefont {Benalcazar}},
  \bibinfo {author} {\bibfnamefont {S.}~\bibnamefont {Huang}}, \bibinfo
  {author} {\bibfnamefont {M.~J.}\ \bibnamefont {Collins}}, \bibinfo {author}
  {\bibfnamefont {K.~P.}\ \bibnamefont {Chen}}, \bibinfo {author}
  {\bibfnamefont {T.~L.}\ \bibnamefont {Hughes}},\ and\ \bibinfo {author}
  {\bibfnamefont {M.~C.}\ \bibnamefont {Rechtsman}},\ }\href
  {https://doi.org/10.1038/s41566-018-0179-3} {\bibfield  {journal} {\bibinfo
  {journal} {Nature Photonics}\ }\textbf {\bibinfo {volume} {12}},\ \bibinfo
  {pages} {408} (\bibinfo {year} {2018})}\BibitemShut {NoStop}%
\bibitem [{\citenamefont {Mittal}\ \emph {et~al.}(2019)\citenamefont {Mittal},
  \citenamefont {Orre}, \citenamefont {Zhu}, \citenamefont {Gorlach},
  \citenamefont {Poddubny},\ and\ \citenamefont {Hafezi}}]{Mittal2019}%
  \BibitemOpen
  \bibfield  {author} {\bibinfo {author} {\bibfnamefont {S.}~\bibnamefont
  {Mittal}}, \bibinfo {author} {\bibfnamefont {V.~V.}\ \bibnamefont {Orre}},
  \bibinfo {author} {\bibfnamefont {G.}~\bibnamefont {Zhu}}, \bibinfo {author}
  {\bibfnamefont {M.~A.}\ \bibnamefont {Gorlach}}, \bibinfo {author}
  {\bibfnamefont {A.}~\bibnamefont {Poddubny}},\ and\ \bibinfo {author}
  {\bibfnamefont {M.}~\bibnamefont {Hafezi}},\ }\href
  {https://doi.org/10.1038/s41566-019-0452-0} {\bibfield  {journal} {\bibinfo
  {journal} {Nature Photonics}\ }\textbf {\bibinfo {volume} {13}},\ \bibinfo
  {pages} {692} (\bibinfo {year} {2019})}\BibitemShut {NoStop}%
\bibitem [{\citenamefont {Ota}\ \emph {et~al.}(2019)\citenamefont {Ota},
  \citenamefont {Liu}, \citenamefont {Katsumi}, \citenamefont {Watanabe},
  \citenamefont {Wakabayashi}, \citenamefont {Arakawa},\ and\ \citenamefont
  {Iwamoto}}]{Ota:19}%
  \BibitemOpen
  \bibfield  {author} {\bibinfo {author} {\bibfnamefont {Y.}~\bibnamefont
  {Ota}}, \bibinfo {author} {\bibfnamefont {F.}~\bibnamefont {Liu}}, \bibinfo
  {author} {\bibfnamefont {R.}~\bibnamefont {Katsumi}}, \bibinfo {author}
  {\bibfnamefont {K.}~\bibnamefont {Watanabe}}, \bibinfo {author}
  {\bibfnamefont {K.}~\bibnamefont {Wakabayashi}}, \bibinfo {author}
  {\bibfnamefont {Y.}~\bibnamefont {Arakawa}},\ and\ \bibinfo {author}
  {\bibfnamefont {S.}~\bibnamefont {Iwamoto}},\ }\href
  {https://doi.org/10.1364/OPTICA.6.000786} {\bibfield  {journal} {\bibinfo
  {journal} {Optica}\ }\textbf {\bibinfo {volume} {6}},\ \bibinfo {pages} {786}
  (\bibinfo {year} {2019})}\BibitemShut {NoStop}%
\bibitem [{\citenamefont {El~Hassan}\ \emph {et~al.}(2019)\citenamefont
  {El~Hassan}, \citenamefont {Kunst}, \citenamefont {Moritz}, \citenamefont
  {Andler}, \citenamefont {Bergholtz},\ and\ \citenamefont
  {Bourennane}}]{ElHassan2019}%
  \BibitemOpen
  \bibfield  {author} {\bibinfo {author} {\bibfnamefont {A.}~\bibnamefont
  {El~Hassan}}, \bibinfo {author} {\bibfnamefont {F.~K.}\ \bibnamefont
  {Kunst}}, \bibinfo {author} {\bibfnamefont {A.}~\bibnamefont {Moritz}},
  \bibinfo {author} {\bibfnamefont {G.}~\bibnamefont {Andler}}, \bibinfo
  {author} {\bibfnamefont {E.~J.}\ \bibnamefont {Bergholtz}},\ and\ \bibinfo
  {author} {\bibfnamefont {M.}~\bibnamefont {Bourennane}},\ }\href
  {https://doi.org/10.1038/s41566-019-0519-y} {\bibfield  {journal} {\bibinfo
  {journal} {Nature Photonics}\ }\textbf {\bibinfo {volume} {13}},\ \bibinfo
  {pages} {697} (\bibinfo {year} {2019})}\BibitemShut {NoStop}%
\bibitem [{\citenamefont {Xue}\ \emph {et~al.}(2019)\citenamefont {Xue},
  \citenamefont {Yang}, \citenamefont {Gao}, \citenamefont {Chong},\ and\
  \citenamefont {Zhang}}]{NatMaterXue2019}%
  \BibitemOpen
  \bibfield  {author} {\bibinfo {author} {\bibfnamefont {H.}~\bibnamefont
  {Xue}}, \bibinfo {author} {\bibfnamefont {Y.}~\bibnamefont {Yang}}, \bibinfo
  {author} {\bibfnamefont {F.}~\bibnamefont {Gao}}, \bibinfo {author}
  {\bibfnamefont {Y.}~\bibnamefont {Chong}},\ and\ \bibinfo {author}
  {\bibfnamefont {B.}~\bibnamefont {Zhang}},\ }\href
  {https://doi.org/10.1038/s41563-018-0251-x} {\bibfield  {journal} {\bibinfo
  {journal} {Nature Materials}\ }\textbf {\bibinfo {volume} {18}},\ \bibinfo
  {pages} {108} (\bibinfo {year} {2019})}\BibitemShut {NoStop}%
\bibitem [{\citenamefont {Ni}\ \emph {et~al.}(2019)\citenamefont {Ni},
  \citenamefont {Weiner}, \citenamefont {Al{\`u}},\ and\ \citenamefont
  {Khanikaev}}]{Ni2019}%
  \BibitemOpen
  \bibfield  {author} {\bibinfo {author} {\bibfnamefont {X.}~\bibnamefont
  {Ni}}, \bibinfo {author} {\bibfnamefont {M.}~\bibnamefont {Weiner}}, \bibinfo
  {author} {\bibfnamefont {A.}~\bibnamefont {Al{\`u}}},\ and\ \bibinfo {author}
  {\bibfnamefont {A.~B.}\ \bibnamefont {Khanikaev}},\ }\href
  {https://doi.org/10.1038/s41563-018-0252-9} {\bibfield  {journal} {\bibinfo
  {journal} {Nature Materials}\ }\textbf {\bibinfo {volume} {18}},\ \bibinfo
  {pages} {113} (\bibinfo {year} {2019})}\BibitemShut {NoStop}%
\bibitem [{\citenamefont {Zhang}\ \emph
  {et~al.}(2019{\natexlab{a}})\citenamefont {Zhang}, \citenamefont {Xie},
  \citenamefont {Wang}, \citenamefont {Xu}, \citenamefont {Tian}, \citenamefont
  {Jiang}, \citenamefont {Lu},\ and\ \citenamefont {Chen}}]{Zhang2019}%
  \BibitemOpen
  \bibfield  {author} {\bibinfo {author} {\bibfnamefont {X.}~\bibnamefont
  {Zhang}}, \bibinfo {author} {\bibfnamefont {B.-Y.}\ \bibnamefont {Xie}},
  \bibinfo {author} {\bibfnamefont {H.-F.}\ \bibnamefont {Wang}}, \bibinfo
  {author} {\bibfnamefont {X.}~\bibnamefont {Xu}}, \bibinfo {author}
  {\bibfnamefont {Y.}~\bibnamefont {Tian}}, \bibinfo {author} {\bibfnamefont
  {J.-H.}\ \bibnamefont {Jiang}}, \bibinfo {author} {\bibfnamefont {M.-H.}\
  \bibnamefont {Lu}},\ and\ \bibinfo {author} {\bibfnamefont {Y.-F.}\
  \bibnamefont {Chen}},\ }\href {https://doi.org/10.1038/s41467-019-13333-9}
  {\bibfield  {journal} {\bibinfo  {journal} {Nature Communications}\ }\textbf
  {\bibinfo {volume} {10}},\ \bibinfo {pages} {5331} (\bibinfo {year}
  {2019}{\natexlab{a}})}\BibitemShut {NoStop}%
\bibitem [{\citenamefont {Zhang}\ \emph
  {et~al.}(2019{\natexlab{b}})\citenamefont {Zhang}, \citenamefont {Wang},
  \citenamefont {Lin}, \citenamefont {Tian}, \citenamefont {Xie}, \citenamefont
  {Lu}, \citenamefont {Chen},\ and\ \citenamefont {Jiang}}]{Zhang2019_2}%
  \BibitemOpen
  \bibfield  {author} {\bibinfo {author} {\bibfnamefont {X.}~\bibnamefont
  {Zhang}}, \bibinfo {author} {\bibfnamefont {H.-X.}\ \bibnamefont {Wang}},
  \bibinfo {author} {\bibfnamefont {Z.-K.}\ \bibnamefont {Lin}}, \bibinfo
  {author} {\bibfnamefont {Y.}~\bibnamefont {Tian}}, \bibinfo {author}
  {\bibfnamefont {B.}~\bibnamefont {Xie}}, \bibinfo {author} {\bibfnamefont
  {M.-H.}\ \bibnamefont {Lu}}, \bibinfo {author} {\bibfnamefont {Y.-F.}\
  \bibnamefont {Chen}},\ and\ \bibinfo {author} {\bibfnamefont {J.-H.}\
  \bibnamefont {Jiang}},\ }\href {https://doi.org/10.1038/s41567-019-0472-1}
  {\bibfield  {journal} {\bibinfo  {journal} {Nature Physics}\ }\textbf
  {\bibinfo {volume} {15}},\ \bibinfo {pages} {582} (\bibinfo {year}
  {2019}{\natexlab{b}})}\BibitemShut {NoStop}%
\bibitem [{\citenamefont {Kempkes}\ \emph {et~al.}(2019)\citenamefont
  {Kempkes}, \citenamefont {Slot}, \citenamefont {van~den Broeke},
  \citenamefont {Capiod}, \citenamefont {Benalcazar}, \citenamefont
  {Vanmaekelbergh}, \citenamefont {Bercioux}, \citenamefont {Swart},\ and\
  \citenamefont {Morais~Smith}}]{Kempkes2019}%
  \BibitemOpen
  \bibfield  {author} {\bibinfo {author} {\bibfnamefont {S.~N.}\ \bibnamefont
  {Kempkes}}, \bibinfo {author} {\bibfnamefont {M.~R.}\ \bibnamefont {Slot}},
  \bibinfo {author} {\bibfnamefont {J.~J.}\ \bibnamefont {van~den Broeke}},
  \bibinfo {author} {\bibfnamefont {P.}~\bibnamefont {Capiod}}, \bibinfo
  {author} {\bibfnamefont {W.~A.}\ \bibnamefont {Benalcazar}}, \bibinfo
  {author} {\bibfnamefont {D.}~\bibnamefont {Vanmaekelbergh}}, \bibinfo
  {author} {\bibfnamefont {D.}~\bibnamefont {Bercioux}}, \bibinfo {author}
  {\bibfnamefont {I.}~\bibnamefont {Swart}},\ and\ \bibinfo {author}
  {\bibfnamefont {C.}~\bibnamefont {Morais~Smith}},\ }\href
  {https://doi.org/10.1038/s41563-019-0483-4} {\bibfield  {journal} {\bibinfo
  {journal} {Nature Materials}\ }\textbf {\bibinfo {volume} {18}},\ \bibinfo
  {pages} {1292} (\bibinfo {year} {2019})}\BibitemShut {NoStop}%
\bibitem [{\citenamefont {Chang}\ and\ \citenamefont
  {Niu}(1996)}]{PhysRevB.53.7010}%
  \BibitemOpen
  \bibfield  {author} {\bibinfo {author} {\bibfnamefont {M.-C.}\ \bibnamefont
  {Chang}}\ and\ \bibinfo {author} {\bibfnamefont {Q.}~\bibnamefont {Niu}},\
  }\href {https://doi.org/10.1103/PhysRevB.53.7010} {\bibfield  {journal}
  {\bibinfo  {journal} {Phys. Rev. B}\ }\textbf {\bibinfo {volume} {53}},\
  \bibinfo {pages} {7010} (\bibinfo {year} {1996})}\BibitemShut {NoStop}%
\bibitem [{\citenamefont {Xiao}\ \emph {et~al.}(2010)\citenamefont {Xiao},
  \citenamefont {Chang},\ and\ \citenamefont {Niu}}]{RevModPhys.82.1959}%
  \BibitemOpen
  \bibfield  {author} {\bibinfo {author} {\bibfnamefont {D.}~\bibnamefont
  {Xiao}}, \bibinfo {author} {\bibfnamefont {M.-C.}\ \bibnamefont {Chang}},\
  and\ \bibinfo {author} {\bibfnamefont {Q.}~\bibnamefont {Niu}},\ }\href
  {https://doi.org/10.1103/RevModPhys.82.1959} {\bibfield  {journal} {\bibinfo
  {journal} {Rev. Mod. Phys.}\ }\textbf {\bibinfo {volume} {82}},\ \bibinfo
  {pages} {1959} (\bibinfo {year} {2010})}\BibitemShut {NoStop}%
\bibitem [{\citenamefont {Atala}\ \emph {et~al.}(2013)\citenamefont {Atala},
  \citenamefont {Aidelsburger}, \citenamefont {Barreiro}, \citenamefont
  {Abanin}, \citenamefont {Kitagawa}, \citenamefont {Demler},\ and\
  \citenamefont {Bloch}}]{Atala2013}%
  \BibitemOpen
  \bibfield  {author} {\bibinfo {author} {\bibfnamefont {M.}~\bibnamefont
  {Atala}}, \bibinfo {author} {\bibfnamefont {M.}~\bibnamefont {Aidelsburger}},
  \bibinfo {author} {\bibfnamefont {J.~T.}\ \bibnamefont {Barreiro}}, \bibinfo
  {author} {\bibfnamefont {D.}~\bibnamefont {Abanin}}, \bibinfo {author}
  {\bibfnamefont {T.}~\bibnamefont {Kitagawa}}, \bibinfo {author}
  {\bibfnamefont {E.}~\bibnamefont {Demler}},\ and\ \bibinfo {author}
  {\bibfnamefont {I.}~\bibnamefont {Bloch}},\ }\href
  {https://doi.org/10.1038/nphys2790} {\bibfield  {journal} {\bibinfo
  {journal} {Nature Physics}\ }\textbf {\bibinfo {volume} {9}},\ \bibinfo
  {pages} {795} (\bibinfo {year} {2013})}\BibitemShut {NoStop}%
\bibitem [{\citenamefont {Mazza}\ \emph {et~al.}(2015)\citenamefont {Mazza},
  \citenamefont {Aidelsburger}, \citenamefont {Tu}, \citenamefont {Goldman},\
  and\ \citenamefont {Burrello}}]{Mazza_2015}%
  \BibitemOpen
  \bibfield  {author} {\bibinfo {author} {\bibfnamefont {L.}~\bibnamefont
  {Mazza}}, \bibinfo {author} {\bibfnamefont {M.}~\bibnamefont {Aidelsburger}},
  \bibinfo {author} {\bibfnamefont {H.-H.}\ \bibnamefont {Tu}}, \bibinfo
  {author} {\bibfnamefont {N.}~\bibnamefont {Goldman}},\ and\ \bibinfo {author}
  {\bibfnamefont {M.}~\bibnamefont {Burrello}},\ }\href
  {https://doi.org/10.1088/1367-2630/17/10/105001} {\bibfield  {journal}
  {\bibinfo  {journal} {New Journal of Physics}\ }\textbf {\bibinfo {volume}
  {17}},\ \bibinfo {pages} {105001} (\bibinfo {year} {2015})}\BibitemShut
  {NoStop}%
\bibitem [{\citenamefont {Meier}\ \emph {et~al.}(2016)\citenamefont {Meier},
  \citenamefont {An},\ and\ \citenamefont {Gadway}}]{Meier2016}%
  \BibitemOpen
  \bibfield  {author} {\bibinfo {author} {\bibfnamefont {E.~J.}\ \bibnamefont
  {Meier}}, \bibinfo {author} {\bibfnamefont {F.~A.}\ \bibnamefont {An}},\ and\
  \bibinfo {author} {\bibfnamefont {B.}~\bibnamefont {Gadway}},\ }\href
  {https://doi.org/10.1038/ncomms13986} {\bibfield  {journal} {\bibinfo
  {journal} {Nature Communications}\ }\textbf {\bibinfo {volume} {7}},\
  \bibinfo {pages} {13986} (\bibinfo {year} {2016})}\BibitemShut {NoStop}%
\bibitem [{\citenamefont {Wang}\ \emph {et~al.}(2017)\citenamefont {Wang},
  \citenamefont {Zhang}, \citenamefont {Chen}, \citenamefont {Yu},\ and\
  \citenamefont {Zhai}}]{PhysRevLett.118.185701}%
  \BibitemOpen
  \bibfield  {author} {\bibinfo {author} {\bibfnamefont {C.}~\bibnamefont
  {Wang}}, \bibinfo {author} {\bibfnamefont {P.}~\bibnamefont {Zhang}},
  \bibinfo {author} {\bibfnamefont {X.}~\bibnamefont {Chen}}, \bibinfo {author}
  {\bibfnamefont {J.}~\bibnamefont {Yu}},\ and\ \bibinfo {author}
  {\bibfnamefont {H.}~\bibnamefont {Zhai}},\ }\href
  {https://doi.org/10.1103/PhysRevLett.118.185701} {\bibfield  {journal}
  {\bibinfo  {journal} {Phys. Rev. Lett.}\ }\textbf {\bibinfo {volume} {118}},\
  \bibinfo {pages} {185701} (\bibinfo {year} {2017})}\BibitemShut {NoStop}%
\bibitem [{\citenamefont {Cardano}\ \emph {et~al.}(2017)\citenamefont
  {Cardano}, \citenamefont {D'Errico}, \citenamefont {Dauphin}, \citenamefont
  {Maffei}, \citenamefont {Piccirillo}, \citenamefont {de~Lisio}, \citenamefont
  {De~Filippis}, \citenamefont {Cataudella}, \citenamefont {Santamato},
  \citenamefont {Marrucci}, \citenamefont {Lewenstein},\ and\ \citenamefont
  {Massignan}}]{Cardano2017}%
  \BibitemOpen
  \bibfield  {author} {\bibinfo {author} {\bibfnamefont {F.}~\bibnamefont
  {Cardano}}, \bibinfo {author} {\bibfnamefont {A.}~\bibnamefont {D'Errico}},
  \bibinfo {author} {\bibfnamefont {A.}~\bibnamefont {Dauphin}}, \bibinfo
  {author} {\bibfnamefont {M.}~\bibnamefont {Maffei}}, \bibinfo {author}
  {\bibfnamefont {B.}~\bibnamefont {Piccirillo}}, \bibinfo {author}
  {\bibfnamefont {C.}~\bibnamefont {de~Lisio}}, \bibinfo {author}
  {\bibfnamefont {G.}~\bibnamefont {De~Filippis}}, \bibinfo {author}
  {\bibfnamefont {V.}~\bibnamefont {Cataudella}}, \bibinfo {author}
  {\bibfnamefont {E.}~\bibnamefont {Santamato}}, \bibinfo {author}
  {\bibfnamefont {L.}~\bibnamefont {Marrucci}}, \bibinfo {author}
  {\bibfnamefont {M.}~\bibnamefont {Lewenstein}},\ and\ \bibinfo {author}
  {\bibfnamefont {P.}~\bibnamefont {Massignan}},\ }\href
  {https://doi.org/10.1038/ncomms15516} {\bibfield  {journal} {\bibinfo
  {journal} {Nature Communications}\ }\textbf {\bibinfo {volume} {8}},\
  \bibinfo {pages} {15516} (\bibinfo {year} {2017})}\BibitemShut {NoStop}%
\bibitem [{\citenamefont {Meier}\ \emph {et~al.}(2018)\citenamefont {Meier},
  \citenamefont {An}, \citenamefont {Dauphin}, \citenamefont {Maffei},
  \citenamefont {Massignan}, \citenamefont {Hughes},\ and\ \citenamefont
  {Gadway}}]{Meiereaat3406}%
  \BibitemOpen
  \bibfield  {author} {\bibinfo {author} {\bibfnamefont {E.~J.}\ \bibnamefont
  {Meier}}, \bibinfo {author} {\bibfnamefont {F.~A.}\ \bibnamefont {An}},
  \bibinfo {author} {\bibfnamefont {A.}~\bibnamefont {Dauphin}}, \bibinfo
  {author} {\bibfnamefont {M.}~\bibnamefont {Maffei}}, \bibinfo {author}
  {\bibfnamefont {P.}~\bibnamefont {Massignan}}, \bibinfo {author}
  {\bibfnamefont {T.~L.}\ \bibnamefont {Hughes}},\ and\ \bibinfo {author}
  {\bibfnamefont {B.}~\bibnamefont {Gadway}}\ }\href
  {https://doi.org/10.1126/science.aat3406} {10.1126/science.aat3406} (\bibinfo
  {year} {2018})\BibitemShut {NoStop}%
\bibitem [{\citenamefont {Maffei}\ \emph {et~al.}(2018)\citenamefont {Maffei},
  \citenamefont {Dauphin}, \citenamefont {Cardano}, \citenamefont
  {Lewenstein},\ and\ \citenamefont {Massignan}}]{Maffei_2018}%
  \BibitemOpen
  \bibfield  {author} {\bibinfo {author} {\bibfnamefont {M.}~\bibnamefont
  {Maffei}}, \bibinfo {author} {\bibfnamefont {A.}~\bibnamefont {Dauphin}},
  \bibinfo {author} {\bibfnamefont {F.}~\bibnamefont {Cardano}}, \bibinfo
  {author} {\bibfnamefont {M.}~\bibnamefont {Lewenstein}},\ and\ \bibinfo
  {author} {\bibfnamefont {P.}~\bibnamefont {Massignan}},\ }\href
  {https://doi.org/10.1088/1367-2630/aa9d4c} {\bibfield  {journal} {\bibinfo
  {journal} {New Journal of Physics}\ }\textbf {\bibinfo {volume} {20}},\
  \bibinfo {pages} {013023} (\bibinfo {year} {2018})}\BibitemShut {NoStop}%
\bibitem [{\citenamefont {Yang}\ \emph {et~al.}(2018)\citenamefont {Yang},
  \citenamefont {Li},\ and\ \citenamefont {Chen}}]{PhysRevB.97.060304}%
  \BibitemOpen
  \bibfield  {author} {\bibinfo {author} {\bibfnamefont {C.}~\bibnamefont
  {Yang}}, \bibinfo {author} {\bibfnamefont {L.}~\bibnamefont {Li}},\ and\
  \bibinfo {author} {\bibfnamefont {S.}~\bibnamefont {Chen}},\ }\href
  {https://doi.org/10.1103/PhysRevB.97.060304} {\bibfield  {journal} {\bibinfo
  {journal} {Phys. Rev. B}\ }\textbf {\bibinfo {volume} {97}},\ \bibinfo
  {pages} {060304} (\bibinfo {year} {2018})}\BibitemShut {NoStop}%
\bibitem [{\citenamefont {Gong}\ and\ \citenamefont
  {Ueda}(2018)}]{PhysRevLett.121.250601}%
  \BibitemOpen
  \bibfield  {author} {\bibinfo {author} {\bibfnamefont {Z.}~\bibnamefont
  {Gong}}\ and\ \bibinfo {author} {\bibfnamefont {M.}~\bibnamefont {Ueda}},\
  }\href {https://doi.org/10.1103/PhysRevLett.121.250601} {\bibfield  {journal}
  {\bibinfo  {journal} {Phys. Rev. Lett.}\ }\textbf {\bibinfo {volume} {121}},\
  \bibinfo {pages} {250601} (\bibinfo {year} {2018})}\BibitemShut {NoStop}%
\bibitem [{\citenamefont {Zhang}\ \emph {et~al.}(2018)\citenamefont {Zhang},
  \citenamefont {Zhang}, \citenamefont {Niu},\ and\ \citenamefont
  {Liu}}]{Zhang2018}%
  \BibitemOpen
  \bibfield  {author} {\bibinfo {author} {\bibfnamefont {L.}~\bibnamefont
  {Zhang}}, \bibinfo {author} {\bibfnamefont {L.}~\bibnamefont {Zhang}},
  \bibinfo {author} {\bibfnamefont {S.}~\bibnamefont {Niu}},\ and\ \bibinfo
  {author} {\bibfnamefont {X.-J.}\ \bibnamefont {Liu}},\ }\href
  {https://doi.org/https://doi.org/10.1016/j.scib.2018.09.018} {\bibfield
  {journal} {\bibinfo  {journal} {Science Bulletin}\ }\textbf {\bibinfo
  {volume} {63}},\ \bibinfo {pages} {1385 } (\bibinfo {year}
  {2018})}\BibitemShut {NoStop}%
\bibitem [{\citenamefont {Zhang}\ \emph
  {et~al.}(2019{\natexlab{c}})\citenamefont {Zhang}, \citenamefont {Zhang},\
  and\ \citenamefont {Liu}}]{PhysRevA.99.053606}%
  \BibitemOpen
  \bibfield  {author} {\bibinfo {author} {\bibfnamefont {L.}~\bibnamefont
  {Zhang}}, \bibinfo {author} {\bibfnamefont {L.}~\bibnamefont {Zhang}},\ and\
  \bibinfo {author} {\bibfnamefont {X.-J.}\ \bibnamefont {Liu}},\ }\href
  {https://doi.org/10.1103/PhysRevA.99.053606} {\bibfield  {journal} {\bibinfo
  {journal} {Phys. Rev. A}\ }\textbf {\bibinfo {volume} {99}},\ \bibinfo
  {pages} {053606} (\bibinfo {year} {2019}{\natexlab{c}})}\BibitemShut
  {NoStop}%
\bibitem [{\citenamefont {Hu}\ and\ \citenamefont
  {Zhao}(2020)}]{PhysRevLett.124.160402}%
  \BibitemOpen
  \bibfield  {author} {\bibinfo {author} {\bibfnamefont {H.}~\bibnamefont
  {Hu}}\ and\ \bibinfo {author} {\bibfnamefont {E.}~\bibnamefont {Zhao}},\
  }\href {https://doi.org/10.1103/PhysRevLett.124.160402} {\bibfield  {journal}
  {\bibinfo  {journal} {Phys. Rev. Lett.}\ }\textbf {\bibinfo {volume} {124}},\
  \bibinfo {pages} {160402} (\bibinfo {year} {2020})}\BibitemShut {NoStop}%
\bibitem [{\citenamefont {Longhi}(2019)}]{Longhi:19}%
  \BibitemOpen
  \bibfield  {author} {\bibinfo {author} {\bibfnamefont {S.}~\bibnamefont
  {Longhi}},\ }\href {https://doi.org/10.1364/OL.44.002530} {\bibfield
  {journal} {\bibinfo  {journal} {Opt. Lett.}\ }\textbf {\bibinfo {volume}
  {44}},\ \bibinfo {pages} {2530} (\bibinfo {year} {2019})}\BibitemShut
  {NoStop}%
\bibitem [{\citenamefont {Qiu}\ \emph {et~al.}(2019)\citenamefont {Qiu},
  \citenamefont {Deng}, \citenamefont {Hu}, \citenamefont {Xue},\ and\
  \citenamefont {Yi}}]{Qiu2019}%
  \BibitemOpen
  \bibfield  {author} {\bibinfo {author} {\bibfnamefont {X.}~\bibnamefont
  {Qiu}}, \bibinfo {author} {\bibfnamefont {T.-S.}\ \bibnamefont {Deng}},
  \bibinfo {author} {\bibfnamefont {Y.}~\bibnamefont {Hu}}, \bibinfo {author}
  {\bibfnamefont {P.}~\bibnamefont {Xue}},\ and\ \bibinfo {author}
  {\bibfnamefont {W.}~\bibnamefont {Yi}},\ }\href
  {https://doi.org/10.1016/j.isci.2019.09.037} {\bibfield  {journal} {\bibinfo
  {journal} {iScience}\ }\textbf {\bibinfo {volume} {20}},\ \bibinfo {pages}
  {392} (\bibinfo {year} {2019})}\BibitemShut {NoStop}%
\bibitem [{\citenamefont {Wang}\ \emph
  {et~al.}(2019{\natexlab{a}})\citenamefont {Wang}, \citenamefont {Qiu},
  \citenamefont {Xiao}, \citenamefont {Zhan}, \citenamefont {Bian},
  \citenamefont {Yi},\ and\ \citenamefont {Xue}}]{Wang2019}%
  \BibitemOpen
  \bibfield  {author} {\bibinfo {author} {\bibfnamefont {K.}~\bibnamefont
  {Wang}}, \bibinfo {author} {\bibfnamefont {X.}~\bibnamefont {Qiu}}, \bibinfo
  {author} {\bibfnamefont {L.}~\bibnamefont {Xiao}}, \bibinfo {author}
  {\bibfnamefont {X.}~\bibnamefont {Zhan}}, \bibinfo {author} {\bibfnamefont
  {Z.}~\bibnamefont {Bian}}, \bibinfo {author} {\bibfnamefont {W.}~\bibnamefont
  {Yi}},\ and\ \bibinfo {author} {\bibfnamefont {P.}~\bibnamefont {Xue}},\
  }\href {https://doi.org/10.1103/PhysRevLett.122.020501} {\bibfield  {journal}
  {\bibinfo  {journal} {Phys. Rev. Lett.}\ }\textbf {\bibinfo {volume} {122}},\
  \bibinfo {pages} {020501} (\bibinfo {year} {2019}{\natexlab{a}})}\BibitemShut
  {NoStop}%
\bibitem [{\citenamefont {Wang}\ \emph
  {et~al.}(2019{\natexlab{b}})\citenamefont {Wang}, \citenamefont {Qiu},
  \citenamefont {Xiao}, \citenamefont {Zhan}, \citenamefont {Bian},
  \citenamefont {Sanders}, \citenamefont {Yi},\ and\ \citenamefont
  {Xue}}]{Wang2019_2}%
  \BibitemOpen
  \bibfield  {author} {\bibinfo {author} {\bibfnamefont {K.}~\bibnamefont
  {Wang}}, \bibinfo {author} {\bibfnamefont {X.}~\bibnamefont {Qiu}}, \bibinfo
  {author} {\bibfnamefont {L.}~\bibnamefont {Xiao}}, \bibinfo {author}
  {\bibfnamefont {X.}~\bibnamefont {Zhan}}, \bibinfo {author} {\bibfnamefont
  {Z.}~\bibnamefont {Bian}}, \bibinfo {author} {\bibfnamefont {B.~C.}\
  \bibnamefont {Sanders}}, \bibinfo {author} {\bibfnamefont {W.}~\bibnamefont
  {Yi}},\ and\ \bibinfo {author} {\bibfnamefont {P.}~\bibnamefont {Xue}},\
  }\href {https://doi.org/10.1038/s41467-019-10252-7} {\bibfield  {journal}
  {\bibinfo  {journal} {Nature Communications}\ }\textbf {\bibinfo {volume}
  {10}},\ \bibinfo {pages} {2293} (\bibinfo {year}
  {2019}{\natexlab{b}})}\BibitemShut {NoStop}%
\bibitem [{\citenamefont {Haller}\ \emph {et~al.}(2020)\citenamefont {Haller},
  \citenamefont {Massignan},\ and\ \citenamefont {Rizzi}}]{Haller2020}%
  \BibitemOpen
  \bibfield  {author} {\bibinfo {author} {\bibfnamefont {A.}~\bibnamefont
  {Haller}}, \bibinfo {author} {\bibfnamefont {P.}~\bibnamefont {Massignan}},\
  and\ \bibinfo {author} {\bibfnamefont {M.}~\bibnamefont {Rizzi}},\ }\href
  {https://doi.org/10.1103/PhysRevResearch.2.033200} {\bibfield  {journal}
  {\bibinfo  {journal} {Phys. Rev. Research}\ }\textbf {\bibinfo {volume}
  {2}},\ \bibinfo {pages} {033200} (\bibinfo {year} {2020})}\BibitemShut
  {NoStop}%
\bibitem [{\citenamefont {Ji}\ \emph {et~al.}(2020)\citenamefont {Ji},
  \citenamefont {Zhang}, \citenamefont {Wang}, \citenamefont {Zhang},
  \citenamefont {Guo}, \citenamefont {Chai}, \citenamefont {Rong},
  \citenamefont {Shi}, \citenamefont {Liu}, \citenamefont {Wang},\ and\
  \citenamefont {Du}}]{Ji2020}%
  \BibitemOpen
  \bibfield  {author} {\bibinfo {author} {\bibfnamefont {W.}~\bibnamefont
  {Ji}}, \bibinfo {author} {\bibfnamefont {L.}~\bibnamefont {Zhang}}, \bibinfo
  {author} {\bibfnamefont {M.}~\bibnamefont {Wang}}, \bibinfo {author}
  {\bibfnamefont {L.}~\bibnamefont {Zhang}}, \bibinfo {author} {\bibfnamefont
  {Y.}~\bibnamefont {Guo}}, \bibinfo {author} {\bibfnamefont {Z.}~\bibnamefont
  {Chai}}, \bibinfo {author} {\bibfnamefont {X.}~\bibnamefont {Rong}}, \bibinfo
  {author} {\bibfnamefont {F.}~\bibnamefont {Shi}}, \bibinfo {author}
  {\bibfnamefont {X.-J.}\ \bibnamefont {Liu}}, \bibinfo {author} {\bibfnamefont
  {Y.}~\bibnamefont {Wang}},\ and\ \bibinfo {author} {\bibfnamefont
  {J.}~\bibnamefont {Du}},\ }\href
  {https://doi.org/10.1103/PhysRevLett.125.020504} {\bibfield  {journal}
  {\bibinfo  {journal} {Phys. Rev. Lett.}\ }\textbf {\bibinfo {volume} {125}},\
  \bibinfo {pages} {020504} (\bibinfo {year} {2020})}\BibitemShut {NoStop}%
\bibitem [{\citenamefont {Wang}\ \emph {et~al.}(2018)\citenamefont {Wang},
  \citenamefont {Wang},\ and\ \citenamefont {Wang}}]{Wang_2018}%
  \BibitemOpen
  \bibfield  {author} {\bibinfo {author} {\bibfnamefont {Q.}~\bibnamefont
  {Wang}}, \bibinfo {author} {\bibfnamefont {D.}~\bibnamefont {Wang}},\ and\
  \bibinfo {author} {\bibfnamefont {Q.-H.}\ \bibnamefont {Wang}},\ }\href
  {https://doi.org/10.1209/0295-5075/124/50005} {\bibfield  {journal} {\bibinfo
   {journal} {{EPL} (Europhysics Letters)}\ }\textbf {\bibinfo {volume}
  {124}},\ \bibinfo {pages} {50005} (\bibinfo {year} {2018})}\BibitemShut
  {NoStop}%
\bibitem [{\citenamefont {Hatsugai}\ and\ \citenamefont
  {Sugi}(2001)}]{doi:10.1142/S0217979201004885}%
  \BibitemOpen
  \bibfield  {author} {\bibinfo {author} {\bibfnamefont {Y.}~\bibnamefont
  {Hatsugai}}\ and\ \bibinfo {author} {\bibfnamefont {A.}~\bibnamefont
  {Sugi}},\ }\href {https://doi.org/10.1142/S0217979201004885} {\bibfield
  {journal} {\bibinfo  {journal} {International Journal of Modern Physics B}\
  }\textbf {\bibinfo {volume} {15}},\ \bibinfo {pages} {2045} (\bibinfo {year}
  {2001})}\BibitemShut {NoStop}%
\bibitem [{\citenamefont {Suzuki}(1990)}]{SUZUKI1990319}%
  \BibitemOpen
  \bibfield  {author} {\bibinfo {author} {\bibfnamefont {M.}~\bibnamefont
  {Suzuki}},\ }\href
  {https://doi.org/https://doi.org/10.1016/0375-9601(90)90962-N} {\bibfield
  {journal} {\bibinfo  {journal} {Physics Letters A}\ }\textbf {\bibinfo
  {volume} {146}},\ \bibinfo {pages} {319 } (\bibinfo {year}
  {1990})}\BibitemShut {NoStop}%
\bibitem [{SM()}]{SM}%
  \BibitemOpen
  \href@noop {} {\ }\bibinfo {note} {See Supplemental Material for a technical
  detail of our numerical calculation, the decoupled cluster argument, the MCQM
  for the different choice of the initial state, the MCQM near the critical
  point, and the results for the bosonic systems, which includes
  Ref~\onlinecite{Su1979}}\BibitemShut {NoStop}%
\bibitem [{\citenamefont {Hatsugai}(2006)}]{doi:10.1143/JPSJ.75.123601}%
  \BibitemOpen
  \bibfield  {author} {\bibinfo {author} {\bibfnamefont {Y.}~\bibnamefont
  {Hatsugai}},\ }\href {https://doi.org/10.1143/JPSJ.75.123601} {\bibfield
  {journal} {\bibinfo  {journal} {Journal of the Physical Society of Japan}\
  }\textbf {\bibinfo {volume} {75}},\ \bibinfo {pages} {123601} (\bibinfo
  {year} {2006})}\BibitemShut {NoStop}%
\bibitem [{\citenamefont {Hatsugai}(2007)}]{Hatsugai_2007}%
  \BibitemOpen
  \bibfield  {author} {\bibinfo {author} {\bibfnamefont {Y.}~\bibnamefont
  {Hatsugai}},\ }\href {https://doi.org/10.1088/0953-8984/19/14/145209}
  {\bibfield  {journal} {\bibinfo  {journal} {Journal of Physics: Condensed
  Matter}\ }\textbf {\bibinfo {volume} {19}},\ \bibinfo {pages} {145209}
  (\bibinfo {year} {2007})}\BibitemShut {NoStop}%
\bibitem [{\citenamefont {Mizoguchi}\ \emph {et~al.}(2019)\citenamefont
  {Mizoguchi}, \citenamefont {Araki},\ and\ \citenamefont
  {Hatsugai}}]{doi:10.7566/JPSJ.88.104703}%
  \BibitemOpen
  \bibfield  {author} {\bibinfo {author} {\bibfnamefont {T.}~\bibnamefont
  {Mizoguchi}}, \bibinfo {author} {\bibfnamefont {H.}~\bibnamefont {Araki}},\
  and\ \bibinfo {author} {\bibfnamefont {Y.}~\bibnamefont {Hatsugai}},\ }\href
  {https://doi.org/10.7566/JPSJ.88.104703} {\bibfield  {journal} {\bibinfo
  {journal} {Journal of the Physical Society of Japan}\ }\textbf {\bibinfo
  {volume} {88}},\ \bibinfo {pages} {104703} (\bibinfo {year}
  {2019})}\BibitemShut {NoStop}%
\end{thebibliography}%
\end{document}


\title{Supplemental Material for ``Detecting Bulk Topology of Quadrupolar Phase from Quench Dynamics"}
\author{Tomonari Mizoguchi}
\affiliation{Department of Physics, University of Tsukuba, Tsukuba, Ibaraki 305-8571, Japan}
\author{Yoshihito Kuno}
\affiliation{Department of Physics, University of Tsukuba, Tsukuba, Ibaraki 305-8571, Japan}
\author{Yasuhiro Hatsugai}
\affiliation{Department of Physics, University of Tsukuba, Tsukuba, Ibaraki 305-8571, Japan}

\maketitle
\section{Division of the Hamiltonian}
To implement the Suzuki-Trotter decomposition described in the main text, 
we decompose the Hamiltonian into five parts:
\begin{eqnarray}
H = H^{(1)} + H^{(2)} + H^{(3)} + H^{(4)} + H^{(5)}.
\end{eqnarray}
Note that $H^{(m)}$'s do not necessarily commute each other. 
Each portion consists of the sum of the local Hamiltonian:
\begin{eqnarray}
H^{(m)} = \sum_{i} h^{(m)}_{i},
\end{eqnarray}
where $h^{(m)}_{i}$'s with the same superscript $(m)$ 
commute each other, i.e., $[h^{(m)}_{i}, h^{(m)}_{j}] = 0$ for $i \neq j$. 

In the present work, we use the following decomposition:
\begin{subequations}
\begin{eqnarray}
H^{(1)} = t_a \sum_{\bm{r}} \left(a^{\dagger}_{\bm{r},1} a_{\bm{r},2} + a^{\dagger}_{\bm{r},4} a_{\bm{r},3}  \right)+ (\mathrm{H.c.}),  
\end{eqnarray}
\begin{eqnarray}
H^{(2)} = t_b \sum_{\bm{r}} \left(a^{\dagger}_{\bm{r},2} a_{\bm{r} + \bm{e}_x,1} + a^{\dagger}_{\bm{r},3} a_{\bm{r}+\bm{e}_x,4}  \right) + (\mathrm{H.c.}),  
\end{eqnarray}
\begin{eqnarray}
H^{(3)} = t_a\sum_{\bm{r}} \left(a^{\dagger}_{\bm{r},1} a_{\bm{r},4} - a^{\dagger}_{\bm{r},2} a_{\bm{r},3}  \right)+ (\mathrm{H.c.}),  
\end{eqnarray}
\begin{eqnarray}
H^{(4)} = t_b  \sum_{\bm{r}} \left(a^{\dagger}_{\bm{r},4} a_{\bm{r} +\bm{e}_y,1}- a^{\dagger}_{\bm{r},3} a_{\bm{r}+\bm{e}_y,2}  \right) + (\mathrm{H.c.}),  
\end{eqnarray}
and 
\begin{eqnarray}
H^{(5)} = H_{\rm int} + H_{\rm rand}.
\end{eqnarray}
\end{subequations}
Note that $\bm{e}_x$ and $\bm{e}_y$ represent unit vectors in $x$ and $y$ directions, respectively,
where the length of the edge of the unit cell is set to be unity.

\section{MCQM in the decoupled cluster limit}
\begin{figure}[!htb]
\begin{center}
\includegraphics[clip,width = 0.75\linewidth]{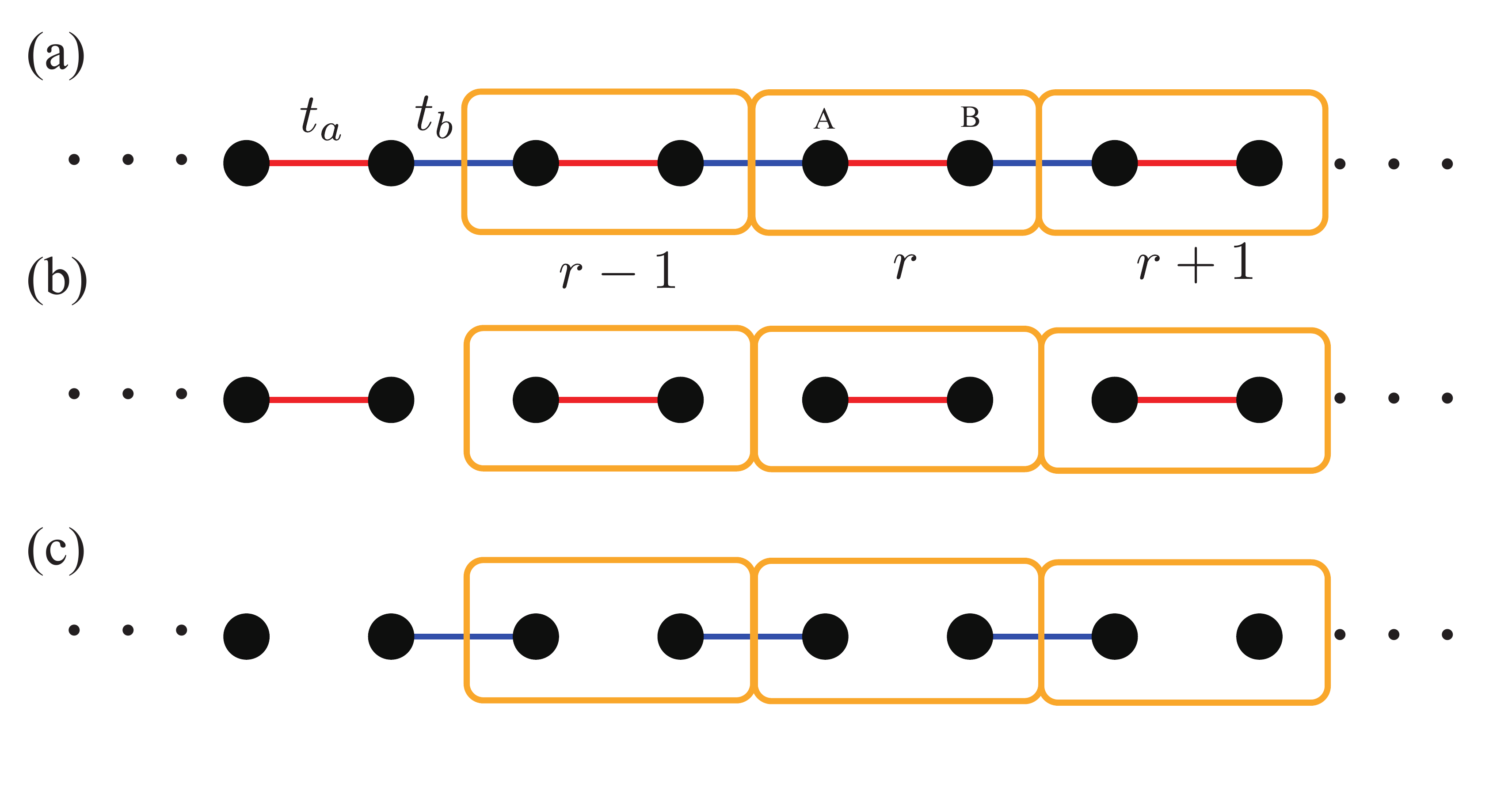}
\vspace{-10pt}
\caption{(a) Schematic figure of 
the SSH model. 
The limits of (b) $t_b = 0$ (trivial)
and (c) $t_a = 0$ (topological).
Orange boxes represent the unit cell.}
\label{fig:ssh}
\end{center}
 \vspace{-10pt}
\end{figure}
In this section, we derive the exact form of the MCQM the decoupled cluster limit for non-interacting case.

\subsection{Warm-up: MCD in the SSH model}
Before discussing the MCQM in the BBH model,
we first discuss the MCD~\cite{Cardano2017,Maffei2018} 
in the Su-Schrieffer-Heeger (SSH) model~\cite{Su1979} in the decoupled dimer limit, 
in order to grasp how the decoupled cluster argument works.
Through this argument, the connection between the MCD and topology is clarified
without relying on the momentum-space picture.

The Hamiltonian of the SSH model reads
\begin{eqnarray}
 H =  t_a  \sum_{r} c^{\dagger}_{r, \mathrm{A}} c_{r, \mathrm{B}} + (\mathrm{H.c.})
 +t_b \sum_{r} c^{\dagger}_{r, \mathrm{B}} c_{r+1, \mathrm{A}} + (\mathrm{H.c.}),
\end{eqnarray}
where $r$ denotes the position of the unit cell and A and B label the sublattices.
The schematic figure of the model is shown in Fig.~\ref{fig:ssh}(a).
When $|t_a| > |t_b|$ ($|t_a|  < |t_b|$), the half-filled ground state is topologically trivial (non-trivial).
This can be evidenced by calculating the topological invariant, i.e., the winding number,
which takes 0 for the trivial case and 1 for the topological case.

Let us consider two different limits, i.e., trivial [Fig.~\ref{fig:ssh}(b)] and non-trivial [Fig.~\ref{fig:ssh}(c)] ones. 
In the following, we assume that the particle is initially localized at A sublattice in the unit cell $r$.
However, the essence of the following argument holds even if the initial position is B sublattice. 
Throughout the unitary time evolution, the particle is completely confined in the dimer 
to which the particle initially belongs;
it is the red bond in the unit $r$ for the trivial case, and the blue bond between the unit cell $r-1$ and $r$ for the non-trivial case.
By solving a two-site problem, one finds $n_{r, \mathrm{A}}(t)  = \cos^2 t_a t$ and $n_{r, \mathrm{B}}(t)  = \sin^2 t_a t$ for the trivial case,
whereas $n_{r, \mathrm{A}}(t)  = \cos^2 t_b t$ and $n_{r-1, \mathrm{B}}(t)  = \sin^2 t_b t$ for the topological case.
Then, the MCD is obtained as $\mathcal{C}^{\rm triv}(t) =r \left(n_{r, \mathrm{A}}(t) -  n_{r, \mathrm{B}}(t) \right) = r \cos 2 t_a t $,
while $\mathcal{C}^{\rm topo}(t) =r n_{r, \mathrm{A}}(t) - (r-1) n_{r, \mathrm{B}}(t) = r \cos 2 t_b t + \sin^2 t_b t$.
Clearly, the long-time average of $\mathcal{C}^{\rm triv}(t)$ is zero, while that of $\mathcal{C}^{\rm topo}(t) $ is $\frac{1}{2}$,
meaning that  the MCD obtained by the momentum-space argument (i.e., the MCD equals to the half of the winding number)
is successfully reproduced.

From the decoupled dimer argument, we obtain the intuitive understanding of the MCD,
namely, the strong inter-unit-cell bond is vital to obtain the finite value.
This intuition is useful for the MCQM in the BBH model, as we will elucidate below.

\subsection{Four-site problem}
\begin{figure}[!htb]
\begin{center}
\includegraphics[clip,width = 0.25\linewidth]{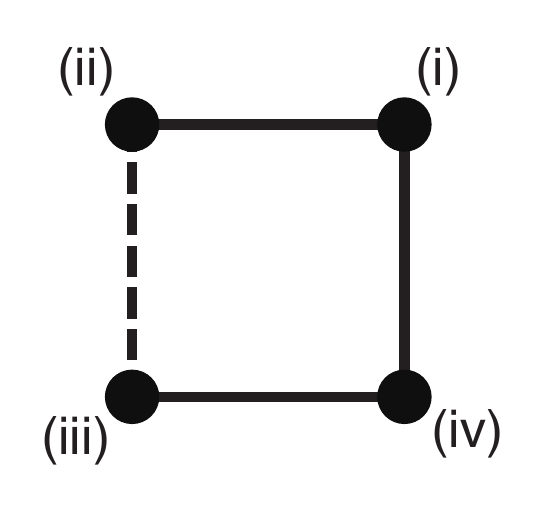}
\vspace{-10pt}
\caption{Schematic figure of the four-site plaquette.}
\label{fig:initial}
\end{center}
 \vspace{-10pt}
\end{figure}
Let us now move on to the BBH model.
To calculate the MCQM for the decoupled cases, we first solve the four-site problem (Fig.~\ref{fig:initial}): 
\begin{eqnarray}
H_{\textrm{4-site}} = \bm{a}^\dagger \mathcal{H} \bm{a},
\end{eqnarray}
where $\bm{a} = \left( a_{\rm (i)}, a_{\rm(ii)}, a_{\rm(iii)}, a_{\rm (iv)} \right)^{\rm T}$
and 
\begin{eqnarray}
\mathcal{H} = \left(
\begin{matrix}
 0 & -1 & 0 & -1\\
 -1 & 0 & 1 & 0 \\
 0 & 1 & 0 & -1 \\
 -1 & 0 & -1 & 0\\
\end{matrix}
\right).
\end{eqnarray}
Note that we set the hopping of the solid (dashed) bonds in Fig.~\ref{fig:initial} as $-1$ ($+1$) for simplicity. 

The eigenstate of $H_{\textrm{4-site}}$, which is denoted by $\gamma_{\xi}^\dagger$ ($\xi = 1,2,3,4$) and satisfies $[\gamma^\dagger_{\xi} , H_{\textrm{4-site}}] =\varepsilon_\xi \gamma^\dagger_{\xi}$, is written as
\begin{eqnarray}
\gamma^\dagger_{\xi}  = \bm{a}^\dagger \cdot \bm{\phi}_{\xi}.
\end{eqnarray}
The eigenenergy $\varepsilon_\xi$ is $-\sqrt{2}$ and $+\sqrt{2}$, for $\xi = 1,2$ and $\xi = 3,4$, respectively.
The wave functions $\bm{\phi}_{\xi}$ are
\begin{subequations}
\begin{eqnarray}
\bm{\phi}_{1} = \left( -\frac{1}{2}, -\frac{1}{\sqrt{2}} , \frac{1}{2}, 0 \right)^{\rm T}, \label{eq:psi1}
\end{eqnarray}
\begin{eqnarray}
\bm{\phi}_{2} = \left(\frac{1}{2}, 0 ,\frac{1}{2}, \frac{1}{\sqrt{2}}\right)^{\rm T},\label{eq:psi2}
\end{eqnarray}
\begin{eqnarray}
\bm{\phi}_{3} = \left(\frac{1}{2}, -\frac{1}{\sqrt{2}} , -\frac{1}{2} , 0 \right)^{\rm T},\label{eq:psi3}
\end{eqnarray}
and 
\begin{eqnarray}
\bm{\phi}_{4} = \left(\frac{1}{2},0 ,\frac{1}{2},  -\frac{1}{\sqrt{2}} \right)^{\rm T}. \label{eq:psi4}
\end{eqnarray}
\end{subequations}

\subsection{Trivial case}
For the trivial case [Fig.~2(a) in the main text], 
two particles are placed on different plaquettes from each other;
one is at the sublattice 3 in the unit cell $(r_x,r_y)$, 
and the other is at the sublattice 1 in $(r_x + 1,r_y + 1)$.
Therefore, we can calculate the contributions from two particles separately,
within the single-particle level.

To proceed, we first derive the explicit form of the time-dependent
particle density of each site for a generic choice of the initial state.
Let $\ket{\psi(0)}$ be the single-particle initial state:
\begin{eqnarray}
\ket{\psi(0)} = \bm{a}^{\dagger} \cdot \bm{\psi}(0) \ket{0},
\end{eqnarray}
with $\ket{0}$ being the vacuum state,
and $\bm{\psi(0)}$ being an generic initial state whose explicit form is
\begin{eqnarray}
\bm{\psi}(0) = \left(\psi_1, \psi_2, \psi_3, \psi_4 \right)^{\rm T}.
\end{eqnarray}
Note that $\bm{\psi}(0)$ is normalized such that $\sum_{\alpha=1}^4 |\psi_\alpha|^2 = 1$. 
Then, one can explicitly calculate 
the state at time $t$, $\ket{\psi(t)}$, 
as 
\begin{eqnarray}
\ket{\psi(t)} &=& 
e^{-i H_{\rm4-site} t} \ket{\psi(0)} \nonumber \\
&=& \sum_{\xi = 1}^{4} e^{- i \varepsilon_{\xi} t} [\bm{\phi}_{\xi} \cdot \bm{\psi}(0)] \gamma_{\xi}^{\dagger} \ket{0}.
\end{eqnarray}
Writing  
\begin{eqnarray}
\ket{\psi(t)} =\bm{a}^\dagger \cdot \bm{\psi}(t) \ket{0},
\end{eqnarray}
and using Eqs.~(\ref{eq:psi1})-(\ref{eq:psi4}), 
one has
\begin{eqnarray} 
\bm{\psi}(t) =
\left(
\begin{array}{c}
\psi_1 \cos \sqrt{2} t +\frac{i}{\sqrt{2}} \sin \sqrt{2} t (\psi_2 + \psi_4) \\
\psi_2 \cos \sqrt{2} t +\frac{i}{\sqrt{2}} \sin \sqrt{2} t (\psi_1 - \psi_3) \\
\psi_3 \cos \sqrt{2} t + \frac{i}{\sqrt{2}} \sin \sqrt{2} t (\psi_4 - \psi_2) \\
\psi_4 \cos \sqrt{2} t +\frac{i}{\sqrt{2}} \sin \sqrt{2} t ( \psi_1 + \psi_3) \\
\end{array}
\right).  \label{eq:psi}
\end{eqnarray}
It follows form Eq.~(\ref{eq:psi}) that the particle density at each site is given as 
\begin{subequations}
\begin{eqnarray}
n_{\rm (i)} (t)&=&|\psi_1|^2 \cos^2 \sqrt{2} t +  \frac{|\psi_2  + \psi_4|^2}{2} \sin^2 \sqrt{2} t  \nonumber \\
&+&\frac{1}{\sqrt{2}} \sin 2\sqrt{2} t \hspace{0.5mm} \mathrm{Im} \left[\psi_1 (\psi_2^\ast + \psi_4^\ast)\right],  \label{eq:n1}
\end{eqnarray}
\begin{eqnarray}
n_{\rm (ii)} (t)
&=&|\psi_2|^2 \cos^2 \sqrt{2} t +  \frac{|\psi_1 - \psi_3|^2}{2} \sin^2 \sqrt{2} t  \nonumber \\
&+&\frac{1}{\sqrt{2}} \sin 2\sqrt{2} t \hspace{0.5mm} \mathrm{Im} \left[\psi_2 (\psi_1^\ast - \psi_3^\ast)\right],\label{eq:n2}
\end{eqnarray}
\begin{eqnarray}
n_{\rm (iii)}(t) &=&|\psi_3|^2 \cos^2 \sqrt{2} t +  \frac{|\psi_4 - \psi_2|^2}{2} \sin^2 \sqrt{2} t  \nonumber \\
&+&\frac{1}{\sqrt{2}} \sin 2\sqrt{2} t \hspace{0.5mm} \mathrm{Im} \left[\psi_3 (\psi_4^\ast - \psi_2^\ast)\right],\label{eq:n3}
\end{eqnarray}
\begin{eqnarray}
n_{\rm (iv)}(t) &=&|\psi_4|^2 \cos^2 \sqrt{2} t +  \frac{|\psi_1 + \psi_3 |^2}{2} \sin^2 \sqrt{2} t  \nonumber \\
&+&\frac{1}{\sqrt{2}} \sin 2\sqrt{2} t \hspace{0.5mm} \mathrm{Im} \left[\psi_4 (\psi_1^\ast +\psi_3^\ast)\right],
\label{eq:n4}
\end{eqnarray}
\end{subequations}

Using (\ref{eq:n1})-(\ref{eq:n4}), one can calculate the MCQM of each particle. 
For the particle at $(r_x,r_y)$, 
we replace the labels of sites in Fig.~\ref{fig:initial} in the following manner:
\begin{itemize}
\item (i) $\rightarrow$ $(r_x,r_y), 1$
\item (ii) $\rightarrow$ $(r_x,r_y), 2$
\item (iii) $\rightarrow$ $(r_x,r_y), 3$
\item (iv) $\rightarrow$ $(r_x,r_y), 4$
\end{itemize}
The corresponding initial state of this plaquette is $\bm{\psi}(0) = (0,0,1,0)$, thus 
the MCQM is given as
\begin{eqnarray}
\mathcal{C}^{\rm triv (1)}_{\rm q}(t) &=&r_x r_y \left(n_{\rm (i)}(t) -n_{\rm (ii)}(t) +n_{\rm (iii)}(t) -n_{\rm (iv)}(t)  \right) \nonumber \\
&=& r_x r_y \cos 2 \sqrt{2} t. \label{eq:cqm_triv}
\end{eqnarray}
The same calculation can be performed for the particle at $(r_x + 1 ,r_y + 1)$, 
where the corresponding initial state is $\bm{\psi}(0) = (1,0,0,0)$, 
and we have
\begin{eqnarray}
\mathcal{C}^{\rm triv (2)}_{\rm q}(t) &=&(r_x  +1) (r_y+1) 
\left(n_{\rm (i)}(t) -n_{\rm (ii)}(t) +n_{\rm (iii)}(t) -n_{\rm (iv)} (t) \right) \nonumber \\
&=& (r_x+1)  (r_y+1)  \cos 2 \sqrt{2} t. \label{eq:cqm_triv}
\end{eqnarray}
Clearly, its long-time average of $\mathcal{C}^{\rm triv (1)}_{\rm q}(t)  + \mathcal{C}^{\rm triv (2)}_{\rm q}(t) $ is zero, since both of these are proportional to $\cos 2 \sqrt{2} t $.

\subsection{Topological case}
For the topological case [Fig.~2(b) in the main text], the plaquette traverses the unit cells.
The correspondence between Fig.~\ref{fig:initial} and Fig.~2(b) in the main text is as follows:
\begin{itemize}
\item (i) $\rightarrow$ $(r_x+1,r_y+1), 1$
\item (ii) $\rightarrow$ $(r_x,r_y+1), 2$
\item (iii) $\rightarrow$ $(r_x,r_y), 3$
\item (iv) $\rightarrow$ $(r_x + 1,r_y), 4$
\end{itemize}

In this case, two particles are confined in the same cluster, so we need to investigate the two-particle dynamics explicitly.
In the absence of the interaction, the two-particle eigenstates can simply be written as $\ket{(\xi,\xi^\prime)}:=\gamma^\dagger_\xi \gamma^\dagger_{\xi^\prime} \ket{0}$ with $\xi < \xi^\prime$, whose eigenenergy is $\varepsilon_{\xi} + \varepsilon_{\xi^\prime}$.
Writing the local basis as
$\ket{\Phi_1}:= a^\dagger_{\rm (i)} a^\dagger_{\rm(ii)} \ket{0}$,
$\ket{\Phi_2}:= a^\dagger_{\rm(i)} a^\dagger_{\rm(iii)} \ket{0}$,
$\ket{\Phi_3}:= a^\dagger_{\rm(i)} a^\dagger_{\rm(iv)} \ket{0}$,
$\ket{\Phi_4}:= a^\dagger_{\rm(ii)} a^\dagger_{\rm(iii)} \ket{0}$,
$\ket{\Phi_5}:= a^\dagger_{\rm(ii)} a^\dagger_{\rm(iv)} \ket{0}$,
and $\ket{\Phi_6}:= a^\dagger_{\rm(iii)} a^\dagger_{\rm(iv)} \ket{0}$,
the explicit forms of the eigenstates are 
\begin{subequations}
\begin{eqnarray}
\ket{(1,2)} =  \frac{1}{2\sqrt{2}} \left( \ket{\Phi_1} - \ket{\Phi_3} -\ket{\Phi_4} + \ket{\Phi_6} \right) -\frac{1}{2} \left(  \ket{\Phi_2} +\ket{\Phi_5}\right), 
\end{eqnarray}
\begin{eqnarray}
\ket{(1,3)} =  \frac{1}{\sqrt{2}}\left( \ket{\Phi_1} + \ket{\Phi_4} \right), 
\end{eqnarray}
\begin{eqnarray}
\ket{(1,4)} =   \frac{1}{2\sqrt{2}} \left( \ket{\Phi_1} + \ket{\Phi_3} -\ket{\Phi_4} - \ket{\Phi_6} \right) - \frac{1}{2} \left(  \ket{\Phi_2} -\ket{\Phi_5}\right), 
\end{eqnarray}
\begin{eqnarray}
\ket{(2,3)} =  \frac{1}{2\sqrt{2}} \left(- \ket{\Phi_1} - \ket{\Phi_3}  + \ket{\Phi_4} + \ket{\Phi_6} \right)  - \frac{1}{2} \left(  \ket{\Phi_2} -\ket{\Phi_5}\right), 
\end{eqnarray}
\begin{eqnarray}
\ket{(2,4)} =  -\frac{1}{\sqrt{2}}\left( \ket{\Phi_3} + \ket{\Phi_6} \right), 
\end{eqnarray}
and
\begin{eqnarray}
\ket{(3,4)} =  \frac{1}{2\sqrt{2}} \left( \ket{\Phi_1} - \ket{\Phi_3} -\ket{\Phi_4} + \ket{\Phi_6} \right)  + \frac{1}{2} \left(  \ket{\Phi_2} +\ket{\Phi_5}\right).
\end{eqnarray}
\end{subequations}

Using these, we can write down the two-particle wave function at time $t$:
\begin{eqnarray}
\ket{\Psi(t)} = \sum_{\xi < \xi^\prime} e^{-i \left(\varepsilon_{\xi} + \varepsilon_{\xi^\prime}\right)t}
\left[ \langle (\xi,\xi^\prime) |\Psi(0)\rangle \right] \ket{(\xi,\xi^\prime)},
\end{eqnarray}
with $\ket{\Psi(0)}$ being the initial state.
Setting $\ket{\Psi(0)} = \ket{\Phi_2}$ [corresponding to Fig.~2(b) in the main text]
and using the site labeling listed in the beginning of this subsection,
we obtain the MCQM:
\begin{eqnarray}
\mathcal{C}_{\rm q} (t) = \bra{\Psi(t)} \mathcal{Q} \ket{\Psi (t)}  
= \frac{1}{2} + \left( \frac{1}{2} + r_x + r_y + 2r_x r_y\right) \cos2 \sqrt{2}t. \label{eq:topo}
\end{eqnarray}
Taking the long-time average, the second term of Eq.~(\ref{eq:topo}) is vanishing
and we obtain $\bar{\mathcal{C}}_{\rm q} (t)  = \frac{1}{2}$. 

\section{MCQM for different initial state}
\begin{figure}[!htb]
\begin{center}
\includegraphics[clip,width = 0.8\linewidth]{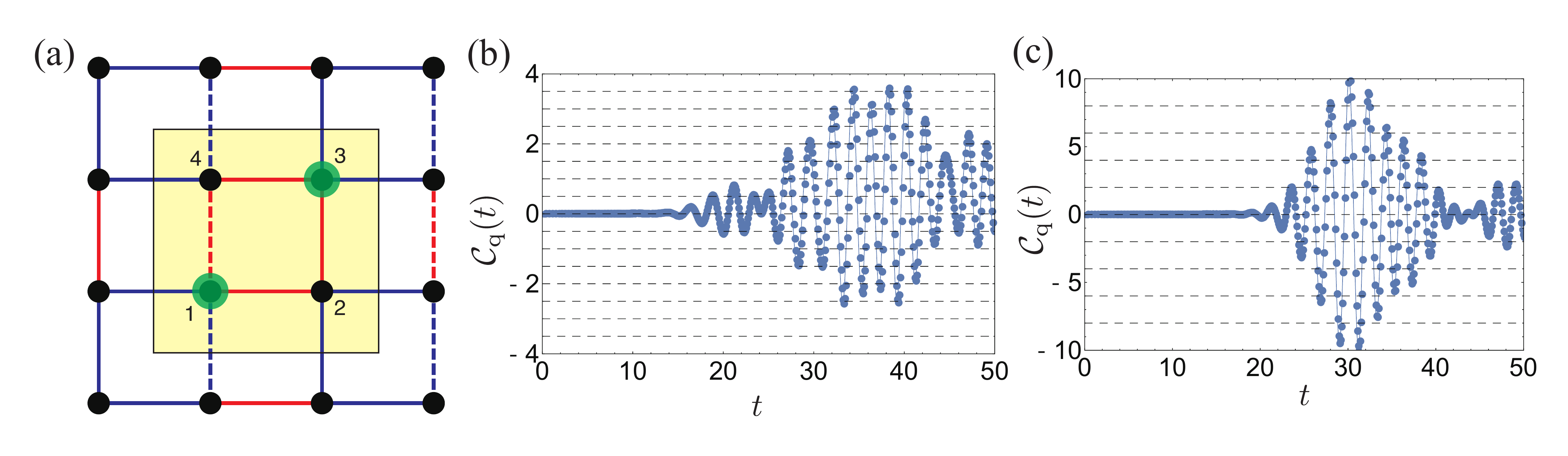}
\vspace{-10pt}
\caption{(a) Schematic figure of the initial state
where two particles are localized at the sublattices 1 and 3 of the same unit cell.
$\mathcal{C}_{\rm q} (t)$ for that case 
with (b) $t_a = -0.3$, $t_b = -1.0$, $V=0$ and (c) $t_a = -1.0$, $t_b=-0.3$, $V=0$.}
  \label{fig:cqm_diff}
 \end{center}
 \vspace{-10pt}
\end{figure}
In this section, we address the sensitivity of the MCQM to the choice of initial state.
The decoupled cluster argument shown in the previous section will predict the MCQM 
in the case where two particles are initially on the same inter-unit-cell plaquette, 
for any combinations of two sublattice indices.
Here, as yet another example, we study the case where the initial positions of two particles are on the same intra-unit-cell plaquette.
Specifically, two particles are initially localized at the sublattice 1 at the unit cell $\left(0,0 \right)$ and the sublattice 3 at the unit cell $\left(0,0\right)$
[Fig.~\ref{fig:cqm_diff}(a)].

The numerical results are shown in Figs.~\ref{fig:cqm_diff}(b) and \ref{fig:cqm_diff}(c). 
We see that the MCQM oscillates around zero in both topological and trivial cases.
This indicates that the initial state with two particles being on the same inter-unit-cell plaquette is essential for 
the distinction from topological and trivial cases using the MCQM.

\section{MCQM near critical point}
\begin{figure}[!htb]
\begin{center}
\includegraphics[clip,width = 0.9\linewidth]{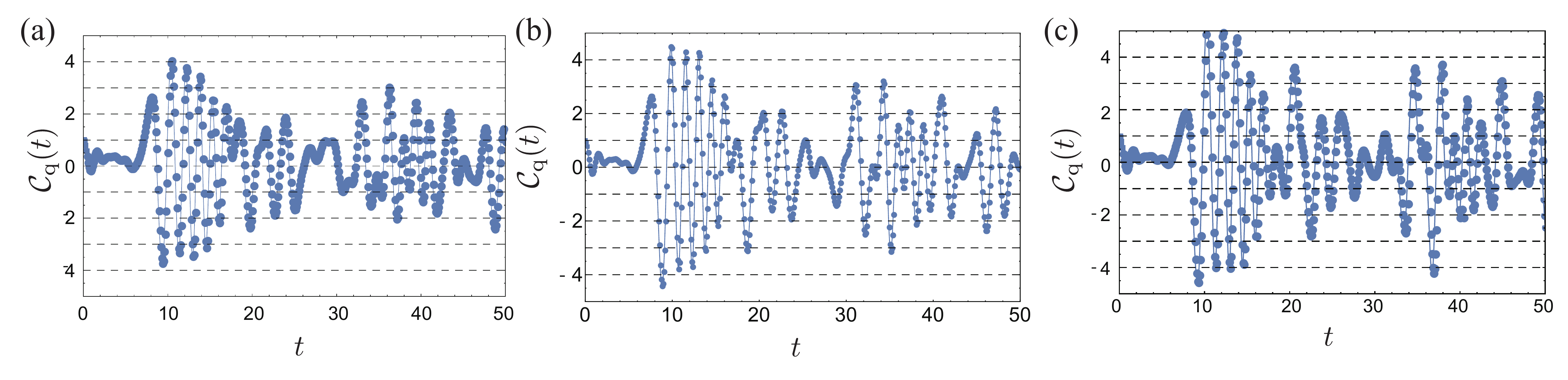}
\vspace{-10pt}
\caption{$\mathcal{C}_{\rm q} (t)$ for (a) $t_a = -0.9$, $t_b = -1.0$, $V=0$ (b) $t_a = -1.0$, $t_b = -1.0$, $V=0$ and (c) $t_a = -1.0$, $t_b = -0.9$, $V=0$. }
  \label{fig:cqm_crit}
 \end{center}
 \vspace{-10pt}
\end{figure}
In this section, we show how the MCQM behaves near the topological transition point $(t_a=t_b)$ for the BBH model.
In Fig.~\ref{fig:cqm_crit}, we plot $\mathcal{C}_{\rm q} (t)$ for 
(a) $t_a = -0.9$, $t_b = -1.0$, $V=0$ (b) $t_a = -1.0$, $t_b = -1.0$, $V=0$ and (c) $t_a = -1.0$, $t_b = -0.9$, $V=0$.
We see that the MCQM does not oscillate around a certain value, which is qualitatively different from the behavior of that 
deep inside the topological or trivial phase shown in Fig.~3 in the main text. 
Therefore, this behavior will serve as a useful hallmark indicating that the system is near the critical point.

\section{Bosonic systems}
\begin{figure}[!htb]
\begin{center}
\includegraphics[clip,width = 0.9\linewidth]{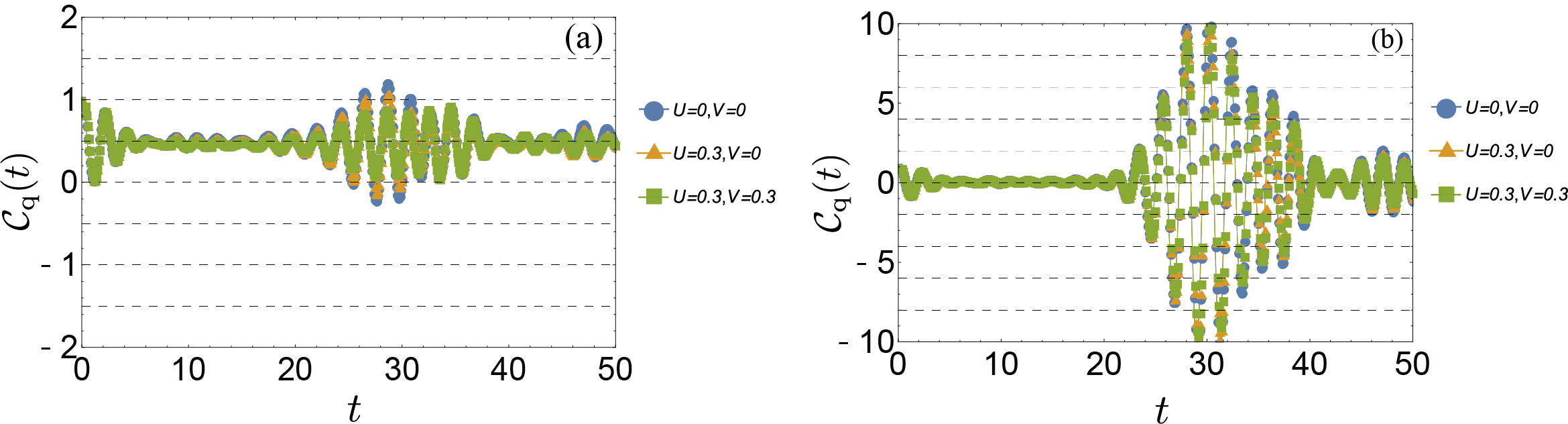}
\vspace{-10pt}
\caption{$\mathcal{C}_{\rm q} (t)$ of the bosonic system without disorders for (a) $t_a = -0.3$, $t_b = -1.0$ and (b) $t_a = -1.0$, $t_b=-0.3$. }
  \label{fig:cqm_b}
 \end{center}
 \vspace{-10pt}
\end{figure}
In this section, we show the results for the bosonic systems.
For bosons, the on-site interaction of the form,
\begin{eqnarray}
H_{\rm on-site} = \frac{U}{2}\sum_i n_i (n_{i}-1),
\end{eqnarray}
is allowed in addition to the terms of Eqs. (1)-(3) in the main text.
Therefore we also consider $H_{\rm on-site}$ in this section. 

\subsection{Numerical results}
In Fig.~\ref{fig:cqm_b}, we plot $\mathcal{C}_{\rm q} (t)$ for the clean system.
We see its behavior is very similar to the fermionic systems;
it oscillates around $1/2$ ($0$) for the topological (trivial) case.
One also see that this behavior is unchanged even if we incorporate small $U$.

\begin{figure}[!htb]
\begin{center}
\includegraphics[clip,width = 0.9\linewidth]{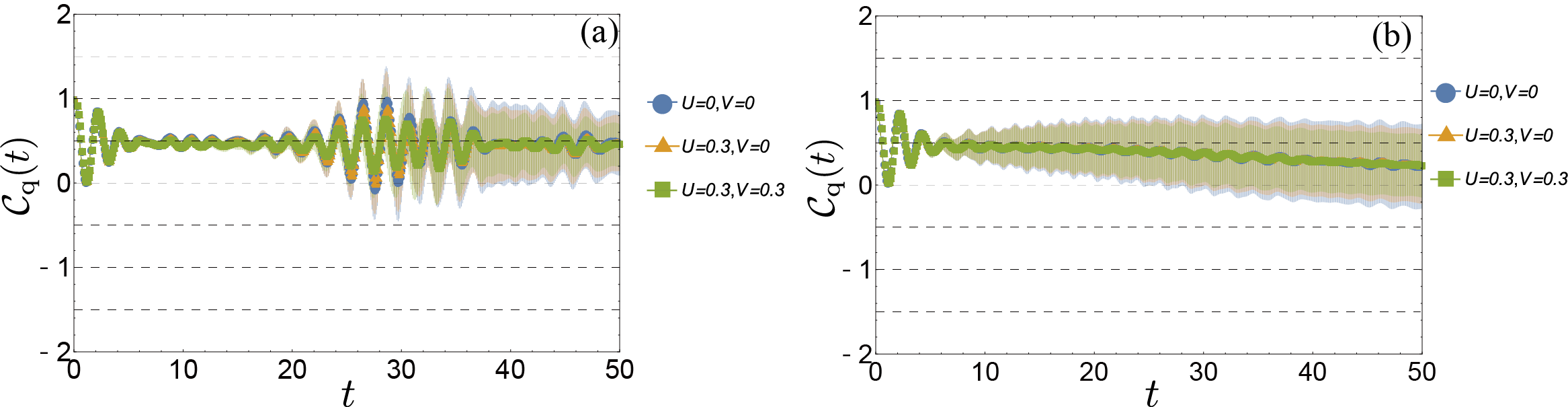}
\vspace{-10pt}
\caption{$\mathcal{C}_{\rm q} (t)$ of the bosonic system with disorders for 
(a) $t_a = -0.3$, $t_b = -1.0$, $W=0.2$ and 
(b) $t_a = -0.3$, $t_b=-1.0$, $W=1.0$. 
The error bars are represented by the shades.}
  \label{fig:cqm_b_dis}
 \end{center}
 \vspace{-10pt}
\end{figure}
In Fig.~\ref{fig:cqm_b_dis}, we plot $\mathcal{C}_{\rm q} (t)$ for the disordered system.
The number of the disorder-potential configurations used to take the average is the same as that for the main text. 
We again see the qualitatively the same behavior as the fermionic case discussed in the main text.

\subsection{Decoupled cluster argument for bosons}
To account for the numerical results shown in the previous subsection,
we again analyze the decoupled cluster limit without interactions. 
Here we focus on the topological case, since the trivial case can be accounted for the single-particle picture
regardless of the particle statistics. 
We note that, even at the non-interacting case, the MCQM for bosons is in general different from that for fermions,
because of the difference of two-particle Hilbert space.

For the four-site problem, the two particle states of bosons 
cab be spanned by the following ortho-normalized states:
$\ket{\Phi^{\rm(B)}_1}:= \frac{1}{\sqrt{2}}a^\dagger_{\rm (i)} a^\dagger_{\rm(i)} \ket{0}$,
$\ket{\Phi^{\rm(B)}_2}:= a^\dagger_{\rm(i)} a^\dagger_{\rm(ii)} \ket{0}$,
$\ket{\Phi^{\rm(B)}_3}:= a^\dagger_{\rm(i)} a^\dagger_{\rm(iii)} \ket{0}$,
$\ket{\Phi^{\rm(B)}_4}:= a^\dagger_{\rm(i)} a^\dagger_{\rm(iv)} \ket{0}$,
$\ket{\Phi^{\rm(B)}_5}:= \frac{1}{\sqrt{2}}a^\dagger_{\rm(ii)} a^\dagger_{\rm(ii)} \ket{0}$,
$\ket{\Phi^{\rm(B)}_6}:= a^\dagger_{\rm(ii)} a^\dagger_{\rm(iii)} \ket{0}$,
$\ket{\Phi^{\rm(B)}_7}:= a^\dagger_{\rm(ii)} a^\dagger_{\rm(iv)} \ket{0}$,
$\ket{\Phi^{\rm(B)}_8}:= \frac{1}{\sqrt{2}}a^\dagger_{\rm(iii)} a^\dagger_{\rm(iii)} \ket{0}$,
$\ket{\Phi^{\rm(B)}_9}:= a^\dagger_{\rm(iii)} a^\dagger_{\rm(iv)} \ket{0}$,
and
$\ket{\Phi^{\rm(B)}_{10}}:= \frac{1}{\sqrt{2}}a^\dagger_{\rm(iv)} a^\dagger_{\rm(iv)} \ket{0}$. 
The eigenstates can be obtained as $\ket{(\xi,\xi^\prime)} = \mathcal{N} \gamma^{\dagger}_\xi \gamma^{\dagger}_{\xi^\prime} \ket{0}$, where $\xi \leq \xi^\prime$ and $\mathcal{N} = \frac{1}{\sqrt{2}}$ for $\xi = \xi^\prime$ and 
$\mathcal{N} = 1$ for  $\xi \neq \xi^\prime$.
Clearly, the eigenenergy of $\ket{(\xi,\xi^\prime)}$ is $\varepsilon_{\xi} + \varepsilon_{\xi^\prime}$. 
For concreteness, we write down all the eigenstates using the basis $\ket{\Phi^{\rm(B)}}$:
\begin{subequations}
\begin{eqnarray}
\ket{(1,1)} = \frac{1}{4} \left( \ket{\Phi^{\rm(B)}_1} +\ket{\Phi^{\rm(B)}_8}\right) 
+ \frac{1}{2} \left(\ket{\Phi^{\rm(B)}_2} + \ket{\Phi^{\rm(B)}_5} -\ket{\Phi^{\rm(B)}_6}  \right)
-\frac{1}{2\sqrt{2}}\ket{\Phi^{\rm(B)}_3},
\end{eqnarray}
\begin{eqnarray} 
\ket{(1,2)} = -\frac{1}{2} \ket{\Phi^{\rm(B)}_7} 
-\frac{1}{2\sqrt{2}} \left(\ket{\Phi^{\rm(B)}_1} + \ket{\Phi^{\rm(B)}_2} + \ket{\Phi^{\rm(B)}_4} + \ket{\Phi^{\rm(B)}_6} - \ket{\Phi^{\rm(B)}_8} - \ket{\Phi^{\rm(B)}_9} \right),
\end{eqnarray}
\begin{eqnarray} 
\ket{(1,3)} = \frac{1}{2} \ket{\Phi^{\rm(B)}_3} 
-\frac{1}{2\sqrt{2}} \left(\ket{\Phi^{\rm(B)}_1} + \ket{\Phi^{\rm(B)}_8} \right) + \frac{1}{\sqrt{2}} \ket{\Phi^{\rm(B)}_5},
\end{eqnarray}
\begin{eqnarray} 
\ket{(1,4)} = \frac{1}{2} \ket{\Phi^{\rm(B)}_7} 
-\frac{1}{2\sqrt{2}} \left(\ket{\Phi^{\rm(B)}_1} + \ket{\Phi^{\rm(B)}_2} - \ket{\Phi^{\rm(B)}_4} + \ket{\Phi^{\rm(B)}_6} - \ket{\Phi^{\rm(B)}_8}  + \ket{\Phi^{\rm(B)}_9}\right),
\end{eqnarray}
\begin{eqnarray} 
\ket{(2,2)} = \frac{1}{4} \left( \ket{\Phi^{\rm(B)}_1} +\ket{\Phi^{\rm(B)}_8}\right) 
+ \frac{1}{2} \left(\ket{\Phi^{\rm(B)}_4} + \ket{\Phi^{\rm(B)}_9}  + \ket{\Phi^{\rm(B)}_{10}}  \right)
 + \frac{1}{2\sqrt{2}}\ket{\Phi^{\rm(B)}_3},
\end{eqnarray}
\begin{eqnarray} 
\ket{(2,3)} = - \frac{1}{2} \ket{\Phi^{\rm(B)}_7} 
+ \frac{1}{2\sqrt{2}} \left(\ket{\Phi^{\rm(B)}_1} - \ket{\Phi^{\rm(B)}_2} + \ket{\Phi^{\rm(B)}_4} - \ket{\Phi^{\rm(B)}_6} - \ket{\Phi^{\rm(B)}_8}  - \ket{\Phi^{\rm(B)}_9}\right),
\end{eqnarray}
\begin{eqnarray} 
\ket{(2,4)} =\frac{1}{2} \ket{\Phi^{\rm(B)}_3} 
+\frac{1}{2\sqrt{2}} \left(\ket{\Phi^{\rm(B)}_1} + \ket{\Phi^{\rm(B)}_8}\right) - \frac{1}{\sqrt{2}}\ket{\Phi^{\rm(B)}_{10}},
\end{eqnarray}
\begin{eqnarray} 
\ket{(3,3)} = \frac{1}{4} \left( \ket{\Phi^{\rm(B)}_1} +\ket{\Phi^{\rm(B)}_8}\right) 
- \frac{1}{2} \left(\ket{\Phi^{\rm(B)}_2} - \ket{\Phi^{\rm(B)}_5} -\ket{\Phi^{\rm(B)}_6} \right)
-\frac{1}{2\sqrt{2}}\ket{\Phi^{\rm(B)}_3},
\end{eqnarray}
\begin{eqnarray} 
\ket{(3,4)} = \frac{1}{2} \ket{\Phi^{\rm(B)}_7} 
+ \frac{1}{2\sqrt{2}} \left(\ket{\Phi^{\rm(B)}_1} - \ket{\Phi^{\rm(B)}_2} - \ket{\Phi^{\rm(B)}_4} 
- \ket{\Phi^{\rm(B)}_6} - \ket{\Phi^{\rm(B)}_8} + \ket{\Phi^{\rm(B)}_9}\right),
\end{eqnarray}
and
\begin{eqnarray} 
\ket{(4,4)} = \frac{1}{4} \left( \ket{\Phi^{\rm(B)}_1} +\ket{\Phi^{\rm(B)}_8}\right) 
- \frac{1}{2} \left(\ket{\Phi^{\rm(B)}_4} + \ket{\Phi^{\rm(B)}_9} -\ket{\Phi^{\rm(B)}_{10}}  \right)
+\frac{1}{2\sqrt{2}}\ket{\Phi^{\rm(B)}_3}.
\end{eqnarray}
\end{subequations}

Using these eigenstates, we can obtain the exact forms of $\ket{\Psi(t)}$ and the MCQM.
For the present choice of the initial state $\ket{\Psi(0)} = \ket{\Phi^{\rm(B)}_3}$, we have 
\begin{eqnarray}
\mathcal{C}_{\rm q} =  \frac{1}{2} + \left(\frac{1}{2} +r_x + r_y + 2r_x r_y  \right)\cos 2 \sqrt{2}t, 
\label{eq:mcqm_boson}
\end{eqnarray}
whose long-time average is $\frac{1}{2}$.
This is exactly the same expression we have for the fermionic case [Eq.~(\ref{eq:topo})],
and also is consistent with the numerical result of Fig.~\ref{fig:cqm_b}(a).